\documentclass[acmsmall]{acmart}
\def\BibTeX{{\rm B\kern-.05em{\sc i\kern-.025em b}\kern-.08emT\kern-.1667em\lower.7ex\hbox{E}\kern-.125emX}}

\usepackage[utf8]{inputenc}
\usepackage{hyperref}
\usepackage{amsmath, amsthm}
\usepackage{amsfonts}
\usepackage{float}
\usepackage{subfig}
\usepackage{algorithm}
\usepackage[noend]{algpseudocode}
\usepackage{tikz} 
\usepackage[shortlabels]{enumitem}
\usetikzlibrary{graphs}
\usetikzlibrary{arrows}
\usetikzlibrary{arrows,positioning,automata,calc}
\usepackage{chngcntr}

\usepackage{lineno}
\theoremstyle{definition}
    \newtheorem{example}{Example}[]
\theoremstyle{definition}
    \newtheorem{theorem}{Theorem}[]

\newcommand{\pair}[1]{\left({#1}\right)}
\newcommand{\sqbra}[1]{\left[{#1}\right]}
\newcommand{\set}[1]{\left\{{#1}\right\}}
\newcommand{\ang}[1]{\left\langle{#1}\right\rangle}
\newcommand{\abs}[1]{\left\lvert{#1}\right\rvert}

\newcommand{\es}{\varnothing}

\newcommand{\nat}{\mathbb{N}}

\begin{document}
	
\title{Who has the last word? Understanding How to Sample Online Discussions}

\author{Gioia Boschi}
\affiliation{%
  \institution{Department of Mathematics, King's College London}
  \streetaddress{Strand, London WC2R 2LS}
  \city{London}
  \country{UK}}
\email{gioia.boschi@kcl.ac.uk}
\orcid{0000-0003-1314-9480}

\author{Anthony P. Young}
\affiliation{%
  \institution{Department of Informatics, King's College London}
  \streetaddress{Bush House, London WC2B 4BG}
  \city{London}
  \country{UK}}
\email{peter.young@kcl.ac.uk}
\orcid{0000-0003-0747-3866}

\author{Sagar Joglekar}
\affiliation{%
  \institution{Department of Informatics, King's College London}
  \streetaddress{Bush House, London WC2B 4BG}
  \city{London}
  \country{UK}}
\email{sagar.joglekar@kcl.ac.uk}

\author{Chiara Cammarota}
  \affiliation{%
    \institution{Department of Physics, Sapienza Università di Roma}
  \streetaddress{P.le A. Moro 5, 00185}
  \city{Rome}
  \country{Italy}}
\affiliation{%
  \institution{Department of Mathematics, King's College London}
  \streetaddress{Strand, London WC2R 2LS}
  \city{London}
  \country{UK}}
\email{chiara.cammarota@kcl.ac.uk}

\author{Nishanth Sastry}
  \affiliation{%
  \institution{Department of Computer Science, University of Surrey}
  \streetaddress{Stag Hill, University Campus, GU2 7XH}
  \city{Guildford}
  \country{UK}}
\affiliation{%
  \institution{Department of Informatics, King's College London}
  \streetaddress{Strand, London WC2R 2LS}
  \city{London}
  \country{UK}}
\email{n.sastry@surrey.ac.uk}

\begin{abstract}

In online debates, as in offline ones, individual utterances or arguments  support or attack each other, leading to some subset of arguments (potentially from different sides of the debate) being considered more relevant than others. However, online conversations are much larger in scale than offline ones, with often hundreds of thousands of users weighing in, collaboratively forming large trees of comments by starting from an original post and replying to each other. In large discussions, readers are often forced to sample a subset of the arguments being put forth. Since such sampling is rarely done in a principled manner, users may not read all the relevant arguments to get a full picture of the debate from a sample. 
This paper is interested in answering the question of how users should sample online conversations to selectively favour the currently justified or accepted positions in the debate.
We apply techniques from argumentation theory and complex networks to build a model that predicts the probabilities of the normatively justified arguments given their location in trees representing idealised online discussions of comments and replies.
Our model shows that the proportion of replies that are supportive, the distribution of the number of replies that comments receive, and the locations of comments that do not receive replies (i.e., the ``leaves'' of the reply tree) all determine the  probability that a comment is a justified argument given its location.
We show that when the distribution of the number of replies is homogeneous along the tree length, for acrimonious discussions (with more attacking comments than supportive ones), the distribution of justified arguments depends on the parity of the tree level which is the distance from the root expressed as number of edges. In supportive discussions, which have more supportive comments than attacks, the probability of having justified comments increases as one moves away from the root. For discussion trees which have a non-homogeneous in-degree distribution, for supportive discussions we observe the same behaviour as before, while for acrimonious discussions we cannot observe the same parity-based distribution.
This is verified with data obtained from the online debating platform Kialo. By predicting the locations of the justified arguments in reply trees, we can therefore suggest which arguments readers should sample, to grasp the currently accepted opinions in such discussions. Our models have important implications for the design of future online debating platforms. 
\end{abstract}


\maketitle

\section{Introduction}\label{sec:intro}

Online discussions have long been an important driver in bringing society and social issues onto the Web, through early platforms such as Usenet and various bulletin board systems in the 1980s \cite{Emerson:83}, and now on social media  on platforms such as Kialo, Reddit, Twitter and Facebook (e.g. \cite{Dekay:12, Gearhart:14, Trilling:15}). As the number of Internet users has grown, so has the scale of the discussions.  For example, the BBC News article reporting on former United Kingdom (UK) Prime Minister Tony Blair's thoughts on Brexit\footnote{\label{fn:Blair}\url{https://www.bbc.co.uk/news/uk-politics-38996179}, last accessed 27/Aug/2020.} has attracted over $10,000$ comments.
Similarly, in the UK, there is an average of 42,600 tweets \emph{per day} exchanged between the Members of Parliament and their followers \cite{Agarwal_Sastry_Wood_2019}, making Twitter the \textit{de facto} platform for digital citizen engagement.

Given the importance of some of the above-mentioned topics, the substantial scale of online discussions creates a problem: Online discussions often contain so many comments that it is  unrealistic to expect a normal Internet user to read every single point being made. Thus, even an interested and impartial reader may only be able to \emph{sample} some of the points in a discussion,  thereby miss crucial points, and end up  making wrong conclusions: A reader viewing a small sample of the whole discussion may be misled into thinking that their favoured arguments are valid in the discussion. However, arguments supporting what a reader considers as an acceptable argument may have been attacked and effectively rebutted in other comments that she was not able to read. Alternatively, views opposing her conclusion may have received important supporting comments which have also been missed by the reader. In either case, the reader has ``missed the big picture'', and come to the wrong conclusions because of sampling a large online discussion. 

This paper seeks to develop better strategies for sampling large online debates by applying the formalism of \textit{bipolar argumentation frameworks} (BAFs) \cite{Cayrol:05}, an object of study in \textit{argumentation theory} \cite{Rahwan:09}, the branch of artificial intelligence concerned with conflict resolution. A BAF is a kind of directed graph (digraph) where the nodes represent \textit{arguments} and each directed edge represents either an \textit{attack} or \textit{support} of one argument towards another. We treat each comment in an online discussion as an argument and represent it as a node. When one comment $a$ replies to another comment $b$, we have an edge from $a$ to $b$, which is either attacking or supporting. In most structured discussion platforms, each comment can only reply to one comment, thereby simplifying the BAF graph to reply \textit{trees}. Our next step is to convert the reply tree BAF into Dung's  \textit{argumentation framework} (AF) \cite{Dung:95}, which provides a  \textit{normative} definition of justified arguments from the attacks: An  argument is considered \textit{justified} if  it is either unrebutted, or every other argument that attacks it is not justified. An argument will be considered \textit{unjustified} if it is attacked by some justified argument. By recursively propagating the justified and unjustified labels across the whole AF, one may identify the subset of \emph{justified} arguments that a reader should focus on given the current set of arguments which have been made.

It is important to note that justified argument can express a range of viewpoints,  potentially from more than one side of a debate, as long as they are not explicitly conflicting. In essence, the unjustified comments (or arguments) are those which have been effectively rebutted, and therefore do not need to be considered by a dispassionate observer of a debate. 
Thus, argumentation theory, and BAF in particular, offers a powerful means to examine online discussions.

Note that the BAF allows us to judge whether an argument is justified or not simply by considering its relation with other arguments -- for example, an argument which is attacked by another justified argument cannot itself be justified. Using a combination of such rules, the BAF allows us to identify justified arguments without having to consider the semantics of the content of individual arguments. In other words,\textit{ once an online discussion has been extracted as the nodes and edges of a BAF, the content does not matter anymore in deciding whether an argument is justified, as all the necessary information is captured in the graph structure}.

Of course, creating the nodes and edges of an argument graph from the natural language of online discussions is non-trivial, and is the subject of an active research area called ``argument mining'' (e.g. see ~\cite{Lippi:16}, \cite{cocarascu2017identifying} and \cite{lawrence2020argument} for  surveys). However, this difficulty is orthogonal to the present work: We consider the reply trees as already formed and ask where in the reply tree (i.e., at which distance from the root or the leaves) can readers find the justified comment. We first answer this question by considering idealised discussions formed from random trees generated by a well-defined in-degree distribution and characterise the patterns of locations where justified comments are clustered in such random trees. Next, we  validate this analysis with data from \textit{Kialo}, an online debating platform \footnote{\url{https://www.kialo.com/}, last accessed 27/Aug/2020.} whose design allows us to straighforwardly extract a BAF: Kialo discussions are well-moderated, such that most comments make a coherent point that is on-topic to the post they are replying to. This allows us to consider each comment as a self-contained and relevant argument that forms a node in the reply tree\footnote{Because of this, we use the terms `comment' and `argument' interchangeably, both for Kialo as well as theoretical results.}. Further, Kialo requires  replies to be classified as pro (support) or con (attack) in relation to the argument they are replying to. Moderation of Kialo discussions also ensures that the 'pro' and 'con' labels are accurate, thus allowing us to reliably label edges between comments and their replies as attacking or supporting. Thus, \textit{the design of Kialo allows us to sidestep problems of mining well-defined arguments from free text}, making it an ideal choice for validating our analytical results. While it is possible to extend our approach to other debate platforms, applying it to other settings where the discussions are not meant to have a logical structure (e.g. Twitter or Reddit), needs further research. We note that argument mining pipelines are starting to be developed for discussions on platforms such as Twitter \cite{Bosc:16}.

\noindent \textbf{Discussion of results}: We use BAF to first investigate a class of idealised discussion trees, in which the in-degree distributions of the nodes, representing the comments, is homogeneous along the length of the tree. With this expression we mean that the degree of a node does not depend on its  \textit{level}, which is its distance from root node (the original post or thesis being debated) measured in number of edges. 
This allows us to calculate the \textit{probability} that an argument is justified as a function of the \textit{level}.
We introduce a parameter $q\in\sqbra{0,1}$, which is the probability that a reply edge is supporting (empirically we measure $\widehat{q}$ as the fraction of supporting edges amongst all edges). Our first result is Theorem \ref{thm:three_behaviours}, which states that in supportive discussions (i.e.,\ reply trees where supportive edges outnumber attacks, giving $q>\frac{1}{2}$), the farther a node is from the root, the higher the probability of it being a justified argument. In acrimonious discussions (reply trees where attacks outnumber edges, i.e. $q<\frac{1}{2}$), the probability of a node being a justified argument depends on the parity of the distance from the leaf levels, and the number of justified comments oscillates from level to level. Lastly, if $q=\frac{1}{2}$, the probability of being justified is independent of the depth, apart from the nodes at the deepest level (which are always justified by default\footnote{In an AF, all unattacked arguments stand as valid. Sec.~\ref{sec:BAFs} gives the theoretical background and  discusses the applicability of this to online discussions, and Sec.~\ref{sec:removing_leaves} discusses alternatives which discount the effect of unreplied arguments in reply trees.}).

Intuitively, in supportive discussions ($ q > 1/2$),  the closer an argument is to the root, the  higher is its chance to have at least one attacking argument in the subtree of replies underneath it, and hence the higher the chance of it being defeated; arguments deeper in the discussion are less likely to be attacked and so ``survive'' the cull for unjustified arguments. Now, consider a different reply tree, which has a chain of comments from a leaf node (which is a justified argument as there is no child node attacking it yet) to a given node, where all comments in the chain are attacking. The leaf node attacks and successfully defeats its parent, which in turn reinstates its grandparent node, thereby defeating the great grandparent, and so on. Thus, in the case where most comments in a reply tree are attacking (i.e., when $q < 1/2 $), arguments are likely to be alternately attacked and reinstated depending on the parity of the distance from the leaf. Finally when $q = 1/2$ the probability of being attacked by a unjustified argument or being supported by a justified one is the same, independently on the level.

We then consider non-homogeneous trees, in which the in-degree of a node depends on its distance from the root. We  show that in this case, when the tree is leaf-heavy, the distribution of justified arguments follows the distributions of the leaves. We will take as example trees where the number of replies follows a scale-free distribution.
We find that Kialo reply trees are well-approximated by non-homogeneous trees, but show some characteristics of homogeneous trees when unrebutted comments are not considered in the count of justified arguments.
Overall, we find that across the models we consider, as well as in the empirical data from Kialo, the leaves of a discussion literally ``have the last word'', i.e. unrebutted arguments at the leaves of reply trees have an enormous influence on the justified arguments: leaves are justified by default and thereby influence which other arguments are justified deeper in the network.  We show that even a conservative position of simply not accepting arguments as justified until they have been supported or attacked by at least one other argument (i.e. only considering non-leaf nodes in the whole reply tree) is not sufficient to remove this influence. We then suggest new methods for calculating the distribution of justified arguments, such that this effect is dampened.

The key contributions of our work can be summarized as follows:

\begin{itemize}
    \item We develop a method (Section \ref{sec:prob_model_intro}) which makes use of argumentation theory (Sections \ref{sec:related_work} and \ref{sec:background}) to calculate the probability of an argument being justified as a function of its level (distance from the root) in discussion trees.
    \item In case of trees with homogeneous in-degree we solve the equations analytically and we identify three regimes of behaviour characterized by the support probability $q$ being smaller, equal or larger than $1/2$ (Section \ref{section:homogeneous}).
    \item We compare the distribution of justified arguments in non-homogeneous reply trees and Kialo graphs, finding both of them strongly dependent on the distribution of leaf nodes in the graph (Section \ref{sec:scale_free_graphs}).
    \item We repeat the analysis removing the leaves from the count of justified arguments and we show that in non-homogeneous trees their contribution to the debate still influences the distribution of justified arguments. In this case for balanced discussions ($q = 1/2$) the probability of an argument being justified can be calculated using only the average number of replies per level. For $q$ smaller or larger than $1/2$ we still need the distribution of leaves per level to fully characterize the probability of an argument being justified (Section \ref{sec:removing_leaves}).
    
\end{itemize}
Our contributions are theoretical in nature, but also provide important insights for the future design of platforms for online discussions. 
The ``takeaway'' for online platforms is that a user should sample  the reply trees at the appropriate distances from the root where the probability of an argument being justified is highest (this sampling can potentially be supported by the platform or its user interface (UI), but can also be done manually by an interested and committed user). The sampling probability is calculated by our model, and depends on factors such as the in-degree distribution of the reply network and the proportion $\widehat{q}$  of replies that are supports.  
The good agreement between synthetic and real data reveals the appropriateness of the use of our probabilistic approach to answer a question that could in general depend on multiple factors of complex human behaviour. 

\section{Related Work}\label{sec:related_work}

As stated in Section \ref{sec:intro}, online discussions cover a vast range of topics and involve many users; this is not surprising due to the growth of access to the Internet, especially through smartphones \cite{owidinternet}. Indeed, 62\% of American adults get their news on social media in 2016 \cite{Allcott:17}, increasing to 67\% in 2017 \cite{pewSurveyNews1}. In the UK, the accounts of UK MPs are collectively being followed by the equivalent of almost 20\% of the UK population \cite{Agarwal:19}. It is reasonable to ask how can we analyse online discussions at scale.
Engaging with large-scale online discussions often lead users to suffer from information overload. For example, UK MPs learn to reply strategically and selectively to citizens concerned with specific topics that are also of interest to the MPs \cite{Agarwal:19}. While MPs are guided by political issues, many discussion platforms have UIs that allow for readers to sort the comments, say from most liked to least liked. This seems to rely on a ``wisdom of the crowds'' effect to have the best points float to the top as indicated by the number of likes, allowing for the user to read the top few points made \cite{ArgSoc:18}. The authors of \cite{Diakopoulos:11} have argued that such comment sorting and structuring mechanisms, including flagging, moderation and ways of detecting relevancy and novelty, can help increase user participation on news comments, improve the quality of comments, and promote constructive discussions. This is what moderation on Kialo also seeks to achieve.\footnote{See \url{https://support.kialo.com/hc/en-us/articles/360000631852-Moderating-Discussions}, last accessed 27/Aug/2020.}
A different perspective on which arguments are more relevant in a discussion is given by \cite{tan2016winning, wei2016post}, where the proposed ranking of comments in forums is based on arguments' persuasiveness.
Sometimes the anonymity and protection of the web can allow people to open up and express their opinions freely. This effect is called the online dis-inhibition effect~\cite{suler2004online} and can be one of the reasons behind the rise of online discussion forums. On the flip-side, guarantee of anonymity could lead to potentially harmful and sometimes toxic behaviour, which was observed in the case studies on Usenet\cite{kayany1998contexts,pfaffenberger1996if}. Such a phenomenon strongly motivates a model of online conversations, which could help sampling the most relevant comments on these forums. 

It is important to study online discussions also because they can affect the offline world. For example, \cite{Dekay:12} has shown that large companies (as defined by Forbes) may actively censor critical comments. However, the magnitude of such effects is open for debate in some cases. For example, \cite{Allcott:17} has clarified the factors behind how much the spread of fake news on social media can be responsible for influencing the 2016 US Presidential Election from the perspective of welfare economics, and argued that there are good reasons to argue that the effects are both small and large.  
But on the individual level, we seek to understand how information can be presented to them such that they read the most compelling of comments made in a discussion.

As mentioned in the introduction, we will model online discussions as directed trees where nodes are arguments and the root is the main thesis. The formation of discussion trees has been widely studied in literature, some examples are \cite{gomez2011modeling, gomez2013likelihood, kumar2010Dynamics, medvedev2019modelling}. In these papers the author propose different models, based on complex networks, to study how discussion cascades form. In particular in \cite{kumar2010Dynamics} the authors propose a model for the formation of the discussion structure based on preferential attachment. We will also use preferential attachment to build our synthetic scale free discussions to mimic Kialo data.

Many examples of the application of complex networks in the study of online discussions are present in the literature. In \cite{StrnadovNeeley2013CharacterizingOD} for example, the authors characterise the discussion forums using network properties. However, they operate on a network of users, which in our case is replaced by a network of arguments. In \cite{Zhang2018Characterizing}, the authors build a computational framework that highlights patterns of interactions in online discussions. They study networks structures to predict the discussion's future, while we use it to understand where the more relevant comments lie. A more applied approach is taken in \cite{konat2016corpus} and \cite{erlin2016social}, where the networks are used respectively in the detection of divisive issues and the measure of the level of participation of students in online platforms.
The comment structure and content is also explored in literature from a pure computational standpoint, to filter out useful feedback on YouTube videos~\cite{Siersdorfer:10}. Another work on comments posted on news articles~\cite{Tsagkias:10} explored the utility of such comments in comparing the engagement potential of news articles. Despite interesting insights from these works, they failed to propose a generalizable approach towards modelling conversations online, not least because of a thematic and qualitative approach.

\subsection{Argumentation Theory in Social Media Analytics}

Given the inevitable diversity of the many views that are expressed and the conflicts that arise, it is important for us to understand how many of these views are consistent, and how various differences can be resolved argumentatively and at scale. This makes online discussions a natural arena for argumentation theory to study. 

Argumentation theory has been applied to both mine structured arguments from natural language text (e.g. \cite{lawrence2020argument, Lippi:16}) and to analyse specific online discussion platforms. For example, \cite{Bosc:16} has designed and tested a pipeline on Twitter. Due to the comparatively noisy messages exchanged on Twitter, the pipeline has to first identify which tweets can be interpreted as self-contained arguments (e.g.  ignoring tweets that consist of just a URL and no other accompanying text), and also infer which replies are attacking or supporting. They then explain how to use an argumentation framework to extract the justified arguments from the discussion. Our work is different and aims to be more general, as we firstly calculate analytically the location of justified arguments in different class of discussion trees, then we apply the same ideas to real online discussions from Kialo. Kialo is a less noisy discussion platform than Twitter --  individual comments in Kialo are moderated to be self-contained arguments and relation between comments is declared. This bypasses the step in \cite{Bosc:16} where tweets need to be identified as self-contained arguments and relations between comments need to be assessed.

Further, \cite{Cabrio:12,Cabrio:13} have applied techniques of argument mining and evaluation to \textit{Debatepedia},\footnote{\url{http://www.debatepedia.org/en/index.php/Welcome_to_Debatepedia\%21}, last accessed 27/Aug/2020.} where attack and support are identified via \textit{textual entailment} (e.g. \cite{dagan2013recognizing}), a technique that aims to reproduce how humans would use common sense to judge whether one piece of text or its negation follow from another piece of text.
In \cite{cocarascu2017identifying} the authors propose an argument mining method to detect attacking and supporting comments in a debate.
Again, given that Kialo already requires users to classify their comments as supporting or attacking, we can bypass such techniques. 
A paper worth mention is also \cite{gao2017random}, which uses graph theory in a theoretical study on argumentation frameworks.

\section{Argumentation Theory and the Kialo Dataset}\label{sec:background}

We now review the relevant technical background in argumentation theory and the procedure by which we have mined data from Kialo.
\subsection{Bipolar Argumentation Theory}\label{sec:BAFs}

\textit{Argumentation theory} is the branch of AI concerned with the rational and transparent resolution of conflicting \textit{arguments}. Arguments and their interactions are represented by \textit{argumentation frameworks} (AFs) \cite{Dung:95}. The type of AFs that we use to represent online discussions are called \textit{bipolar argumentation frameworks} (BAFs) \cite{Cayrol:05}. Formally, a BAF is a structure $\ang{A,R_{sup},R_{att}}$, where $A$ is our \text{set of arguments} and $R_{sup},R_{att}\subseteq A^2$ are binary relations on $A$ that respectively represent supporting and attacking replies, i.e. for $a,b\in A$, $(a,b)\in R_{sup}$ means $a$ \textit{supports} (agrees with) $b$, and $(a,b)\in R_{att}$ means $a$ \textit{attacks} (disagrees with) $b$. We require that $R_{sup}\cap R_{att}=\es$. One can therefore think of $\ang{A,R_{sup},R_{att}}$ as a directed graph (digraph) where supporting (dotted) edges are green and attacking (solid) edges are red.

\begin{example}\label{eg:baf}
Illustrated in Figure \ref{fig:tree2} is the BAF where $A=\set{a,b,c,d,e}$, $R_{att}=\set{(c,b)}$ and $R_{sup}=\set{(d,c),(e,b),(b,a)}$.

\begin{figure}[H]
\begin{center}
\begin{tikzpicture}[>=stealth',shorten >=1pt,node distance=2cm,on grid,initial/.style    ={}]
\tikzset{mystyle/.style={->,relative=false,in=0,out=0}};
\node[state] (a) at (0,0) {$a$};
\node[state] (b) at (2,0) {$b$};
\node[state] (c) at (4,0) {$c$};
\node[state] (d) at (6,0) {$d$};
\node[state] (e) at (2,-1.5) {$e$};
\draw [->, green, thick, dotted] (e) to (b);
\draw [->, green, thick, dotted] (d) to (c);
\draw [->, red, thick] (c) to (b);
\draw [->, green, thick, dotted] (b) to (a);
\end{tikzpicture}
\caption{The BAF from Example \ref{eg:baf}, where green (dotted) edges denote supports and red (solid) edges denote attacks}\label{fig:tree2}
\end{center}
\vspace{-0.5cm}
\end{figure}
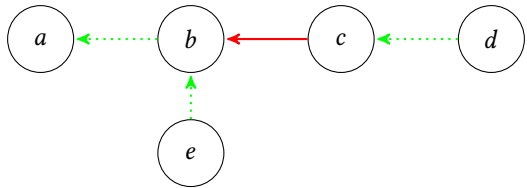
\end{example}

How can we determine the justified arguments in a BAF? Following \cite{Cayrol:05}, we combine supports into attacks, which result in \textit{defeats}; arguments are justified if either they are not defeated, or are \textit{reinstated} by having all their defeaters in turn defeated. More formally, given a BAF $\ang{A,R_{sup},R_{att}}$ and $a,b\in A$, we say a \textit{support path} is a path in the underlying digraph that only traverses support edges. Let $a\to_{sup}b$ denote that there exists a support path from $a$ to $b$. We say $a$ \textit{support-defeats} $b$ iff $\pair{\exists c\in A}\sqbra{a\to_{sup} c,\:R_{att}(c,b)}$, and $a$ \textit{indirectly defeats} $b$ iff $\pair{\exists c\in A}\sqbra{R_{att}(a,c),\:c\to_{sup} b}$. We define the \textit{argumentation framework of $\ang{A,R_{sup},R_{att}}$} to be the digraph $\ang{A,R}$ where $(a,b)\in R$ iff either $a$ support-defeats $b$ or $a$ indirectly defeats $b$ \cite{Cayrol:05, Dung:95, Young:18}. We say $a$ \textit{defeats} $b$ iff $(a,b)\in R$.

\begin{example}\label{eg:baf_to_af}
(Example \ref{eg:baf} continued) The support paths of length 1 in this BAF are $(e,b)$, $(b,a)$ and $(d,c)$, and the support paths of length 2 in this BAF consist of only $(e,b,a)$. Therefore, $e\to_{sup} b$, $b\to_{sup} a$, $d\to_{sup} c$ and $e\to_{sup} a$. As $c$ attacks $b$, we can see that $c$ indirectly defeats $a$, and $d$ support-defeats $b$. Therefore, the corresponding AF of this BAF has the same arguments, but the defeat relation is $R=\set{(c,b),(c,a),(d,b)}$. This is illustrated in Figure \ref{fig:tree3}.

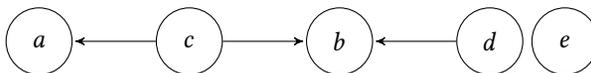
\begin{figure}[H]
\begin{center}
\begin{tikzpicture}[>=stealth',shorten >=1pt,node distance=2cm,on grid,initial/.style    ={}]
\tikzset{mystyle/.style={->,relative=false,in=0,out=0}};
\node[state] (a) at (0,0) {$a$};
\node[state] (c) at (2,0) {$c$};
\node[state] (b) at (4,0) {$b$};
\node[state] (d) at (6,0) {$d$};
\node[state] (e) at (7,0) {$e$};
\draw [->] (c) to (a);
\draw [->] (c) to (b);
\draw [->] (d) to (b);
\end{tikzpicture}
\caption{The corresponding AF of the BAF in Figure \ref{fig:tree2}, from Example \ref{eg:baf_to_af}; argument $e$ is an isolated node.}\label{fig:tree3}
\end{center}

\end{figure}
\end{example}

\noindent If $\ang{A,R_{sup},R_{att}}$ is a tree, then $\ang{A,R}$ is a directed acyclic graph, and we can use Algorithm \ref{alg:grounded} to calculate the set of justified arguments, also called \textit{the grounded extension}.\footnote{This set exists and is unique \cite[Theorem 30]{Dung:95}. This algorithm is a special case of the general definitions of justified arguments \cite{Dung:95}, which also apply to, e.g. cyclic or infinite argumentation frameworks.}

\begin{algorithm}[ht]
\begin{algorithmic}[1]
\Function{GroundedExtension}{$\ang{A,R}$}
  \State $in\gets\es$
  \State $out\gets\es$
  \While {$in\neq A$} \do\\
    \State $in\gets\set{a\in A\:\vline\:\pair{\forall b\in A}(b,a)\notin R}$
    \State $out\gets\set{a\in A\:\vline\:\pair{\exists b\in in}(b,a)\in R}$
    \State $A\gets A-out$
    \State $R\gets R\cap A^2$
  \EndWhile
  \Return $A$
\EndFunction
\end{algorithmic}
\caption{Algorithm for calculating the set of justified arguments of $\ang{A,R}$ (from \cite{Wooldridge:09}).}
\label{alg:grounded}
\end{algorithm}

Intuitively, the algorithm first labels all unattacked arguments as $in$ (justified) and all arguments attacked by the unattacked arguments as $out$ (unjustified). In the context of reply trees, the unattacked arguments correspond to the leaves. The algorithm then excludes the unjustified arguments from the arguments under consideration and consider the next set of unattacked arguments and the arguments attacked by those unattacked arguments... etc. until all arguments are labelled by either $in$ or $out$, which is possible for reply trees.\footnote{Non-tree argumentation frameworks can have arguments that are neither $in$ nor $out$, but that will take us beyond the scope of this paper.}
It has been shown that this algorithm runs in polynomial time \cite{Dunne:09}.

\begin{example}
(Example \ref{eg:baf_to_af} continued) We can apply the algorithm to Figure \ref{fig:tree3}. It is easy to see that after one iteration of the while loop, we get $in=\set{e,d,c}$ and $out=\set{a,b}$. This means $A=in$, so the algorithm halts and returns $\set{e,d,c}$, which is the set of justified arguments of Figure \ref{fig:tree3} and hence of Figure \ref{fig:tree2}.
\end{example}

One criticism of adopting BAFs as a model for online discussions is that its interpretation of support is very strong, akin to logical necessity \cite{Cayrol:13}, i.e. if $c\to_{sup} b$ then $c$ is necessary for $b$, hence any attacker of $c$ must indirectly defeat $b$. This may not hold for the informal logic employed in online discussions, but we allow for this assumption here as we would like to make the supporting arguments as vulnerable to defeat as possible such that only the strongest such arguments survive, i.e. arguments where all their supporters are either undefeated or reinstated. This ``skeptical'' stance seems suitable given that online discussions take place in low-trust environments given its potential to become rife with misinformation. However, even if the discussion environment is well-moderated and seeks to promote good debating practice, one should adopt such a skeptical stance to be able to claim that arguments are justified if they are consistent with defensible foundations.

Another criticism is that unrebutted arguments should not be justified in the context of reply trees. Originally, this is meant to capture two principles - that everything relevant is already represented in the digraph of arguments, and that arguments are assumed to be acceptable by default until shown otherwise - a form of ``lazy'' reasoning \cite{Dung:95}. But in reply trees, the leaf arguments do not have to be justified, especially if the conversation could have degenerated to insults the further it departs from the starting claims. How can we let such claims being justified? We clarify that the term ``justified'' does not denote truth and unrebutted claims are vacuously justified at that point in time until explicitly rebutted by a comment which is made at a later time point. Further, Kialo's moderation policy would not allow debates to degenerate into trolling or insults, so by using Kialo we cannot count those as reasons for why leaf arguments should not being justified. However, we will attempt to address the deficiencies of letting leaves being justified in Section \ref{sec:removing_leaves}.

In summary, we have defined BAFs and how to resolve the conflicts therein through transforming BAFs into AFs and calculating the justified arguments. This follows various normative principles such as that arguments that are neither attacked nor supported are always justified, and two arguments defeating each other cannot be simultaneously justified.

\subsection{Details of the Kialo Dataset}\label{sec:Kialo}

\begin{figure}
\centering
\includegraphics[scale=0.3]{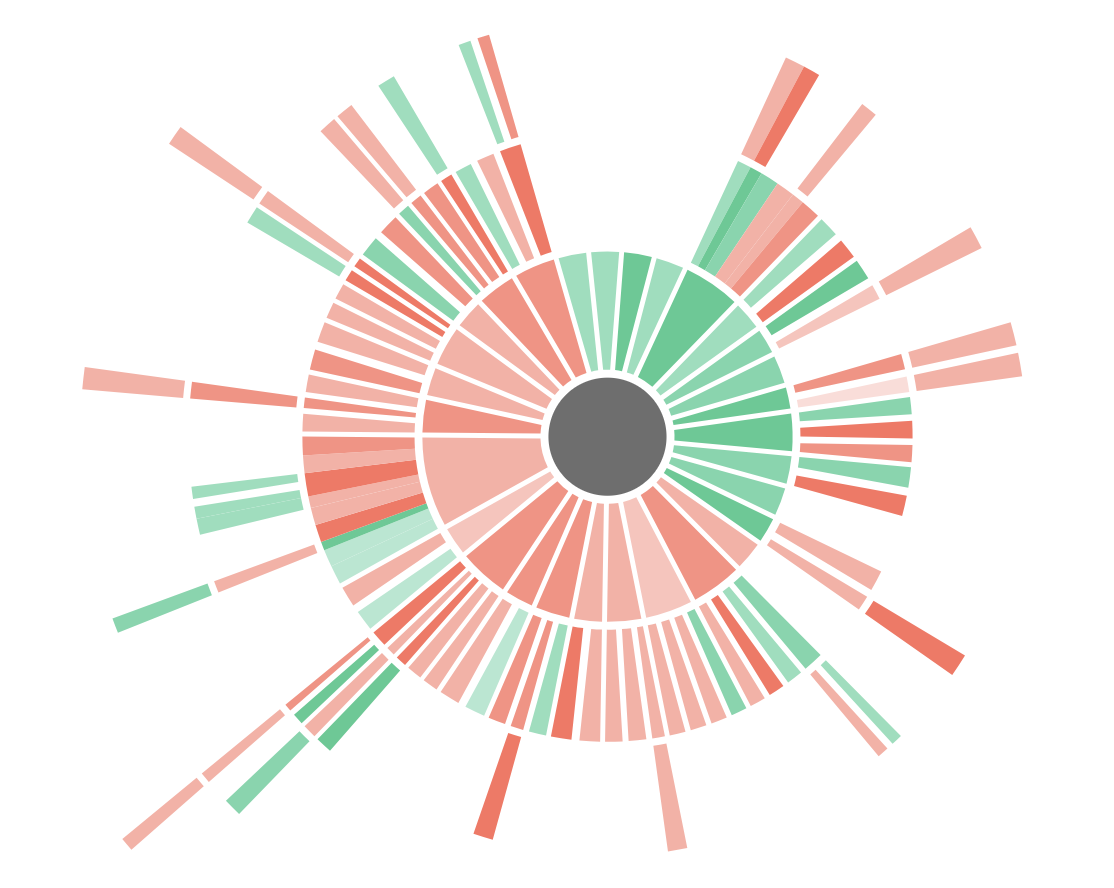}
\includegraphics[scale=0.25]{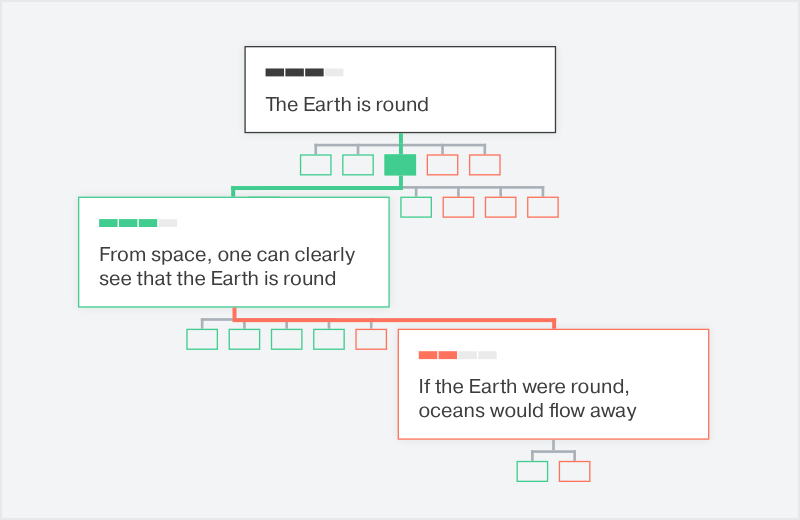}
\caption{Example of a Kialo discussion. The thesis is represented by the grey rectangle on the right diagram and the grey circle on the left diagram. In the left diagram, the concentric annuli represent the replies made at different distance from the root, where each slice of the annulus denotes a single replying comment. Replies are coloured green if they support and red if they attack the previous comment, the shade of the colouring representing the number of likes to the comment, with higher number of likes denoted by a darker shade.}\label{fig:Kialo_example}
\end{figure}

In order to validate our model we use data from discussions hosted on \textit{Kialo}, an online debating platform.\footnote{\url{https://www.kialo.com/}, last accessed 27/Aug/2020.} Figure \ref{fig:Kialo_example} illustrates an example Kialo discussion.\footnote{The left sub-figure is taken from \url{https://stackoverflow.com/questions/49854754/kialo-how-can-i-view-the-argument-topology-map-after-i-have-entered-an-argument}, last accessed 27/Aug/2020. The right sub-figure is taken from \url{http://mycareacademy.org/all/a-new-digital-debating-tool-for-collaborators-kialo/}, last accessed 27/Aug/2020.
} In this section, we outline how discussions are initiated in Kialo and the procedure by which we scraped and cleaned the discussions. 

\subsubsection{Discussions and Sub-discussions on Kialo}\label{sec:Kialo_discussions}

To start a discussion, the user creates a \textit{thesis} along with a \textit{tag} that indexes the discussion. A thesis can have many tags, which increases its visibility to users. Additionally, a discussion can be created with an option to add multiple theses to debate. For example, a discussion could start with an open question like ``Who is the ultimate fighting hero from any fandom?''\footnote{See \url{https://www.kialo.com/who-is-the-ultimate-fighting-hero-from-any-fandom-8857}, last accessed 27/Aug/2020.} and several theses could be proposed as debatable options under this overarching question. In such a situation, this one discussion thesis could spawn multiple sub-discussions, each proposing a candidate fighting hero  (e.g. Superman and Batman), which will give rise to a separate reply tree of their own. 

\subsubsection{Scraping and Cleaning Kialo Discussions}
\begin{figure}
\centering
\includegraphics[width=0.6\columnwidth]{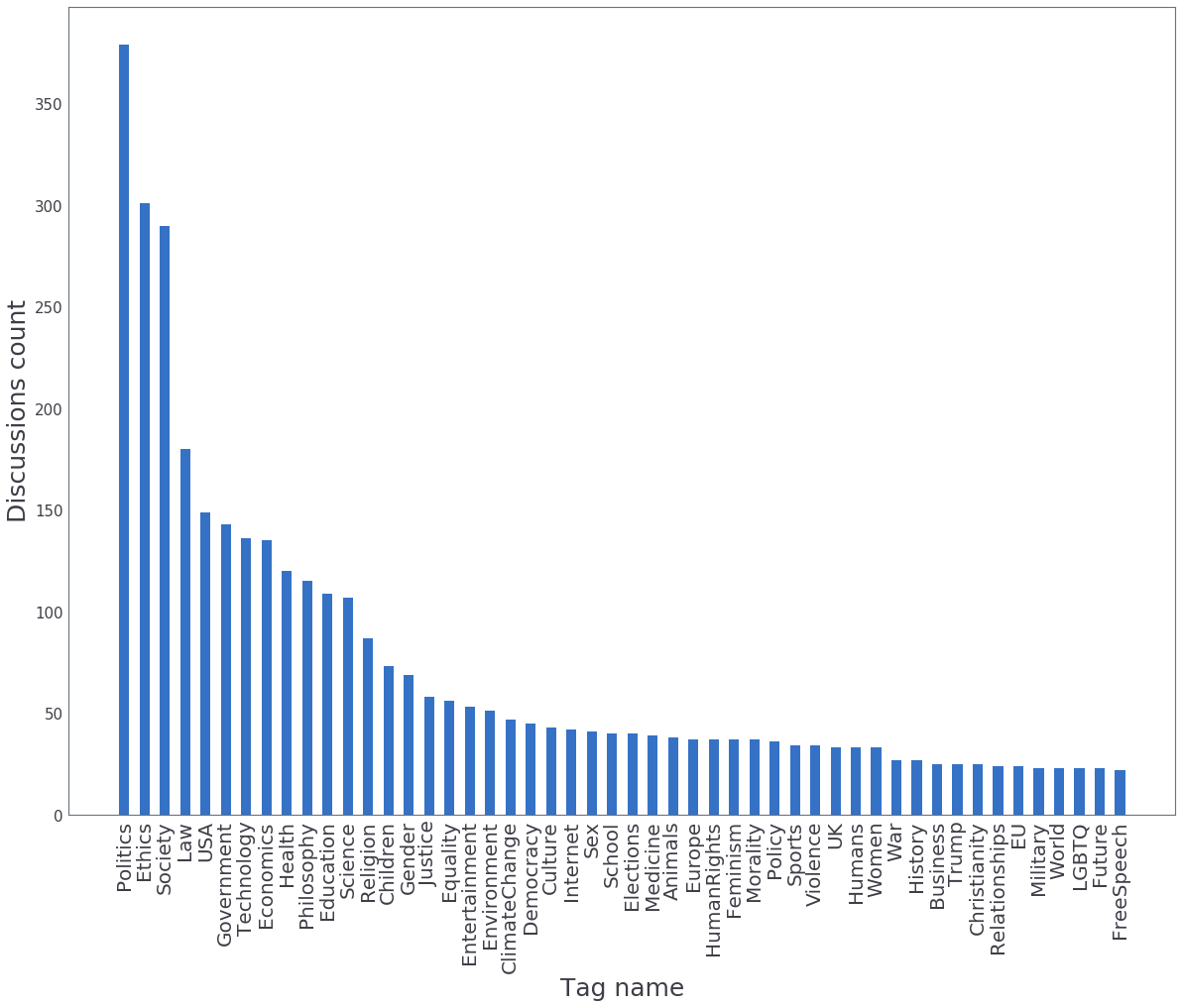}
\caption{ Occurrence of the top fifty topic tags across Kialo discussions. The top five are: politics, ethics, society, law and government. For example, there are just under 300 discussions for politics, and around 150 discussions for law.} \label{fig:tags_example}
\end{figure}

To obtain the dataset, we reverse engineered the Kialo app API, which obtains all the available tags on the Kialo website. This is done by first bootstrapping the query with certain featured tags on Kialo\footnote{See \url{https://www.kialo.com/explore/featured}, last accessed 27/Aug/2019.} and then progressively expanding the tags dataset by adding the co-occuring tags with the bootstrap set. At the end of the process, we were able to get 1120 tags, which covers almost all of the discussions hosted on Kialo as of 28th of January 2020. To verify this claim, we scripted another utility that exploited Kialo's scrolling API to go as far back in the list as possible to get the oldest thread, and we ended up with the same number of threads to view. Figure \ref{fig:tags_example} shows a histogram of the top fifty popular tags among the 1120 along with corresponding number of threads associated with a particular tag.

As the next step we obtained all the discussion threads associated with each of these 1120 tags. This was done by mimicking the tag-based search feature of Kialo and getting all the results that show up for a particular tag based search. As mentioned before, each thread can be associated to one or more tags. Through our data collection scripts, we are able to obtain 1560 discussion threads. Our manual verification gives us a high degree of confidence that this is almost all of the debate activity on the service. We progressively crawl each discussion thread to acquire the data about the tree structure, votes on each argument and the argument text. This also includes all the sub discussion trees resulting due to debates having multiple thesis, as described in Section \ref{sec:Kialo_discussions}. Before analysing the data, we cleaned them by removing all the trees with less than twenty nodes and removing all the discussions with comments that have empty text or deleted branches. We are left with a total 1511 final trees to analyse. 

We also acquire other supplemental meta-data such as the time of posting, the time of editing (if any) and the author meta-data. To our knowledge, this is the most complete snapshot of Kialo, as of 28th of January 2020.\footnote{To aid reproducibility and encourage follow-on work, our data will be shared upon request with the wider research community post publication.} 

All discussions that were crawled from Kialo have a tree structure with a root node that represents the main thesis and each other node is a reply to its parent. Each reply answers only to the argument of the parent, so an answer in favour to a node, does not necessarily  represent a support to the main thesis.
We will represent Kialo discussions as directed trees, where the directed edges point in the direction of the reply. Each edge can have a positive or negative sign, respectively indicating support or attack; the representation of such discussions as bipolar argumentation frameworks is therefore straightforward (Section \ref{sec:BAFs}), and we have ready-made techniques for evaluating which arguments are justified (Section \ref{sec:related_work}).

\begin{figure*}
    \centering
     \subfloat[\label{subfig-1-hist_q}]{\includegraphics[scale=0.27]{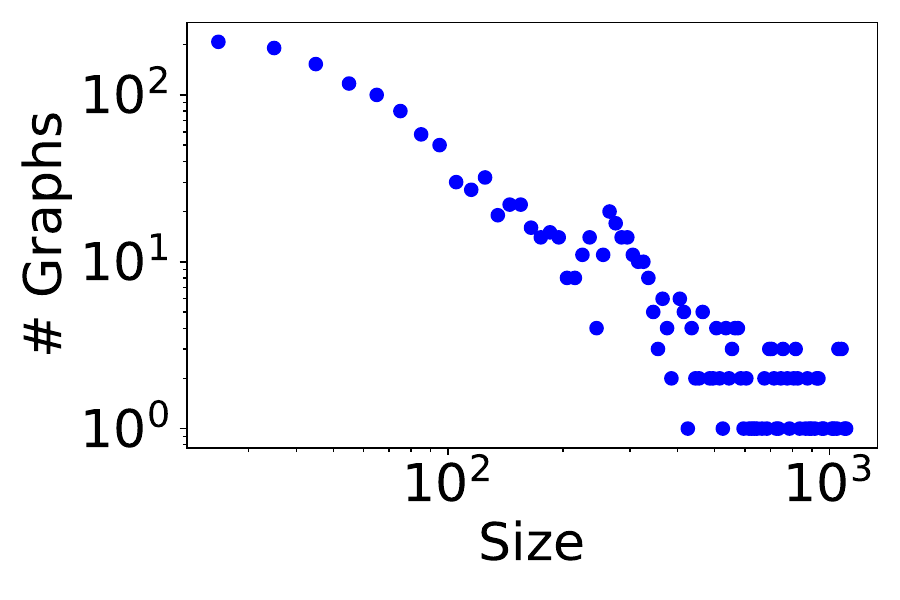}}
    \subfloat[\label{subfig-2-hist_q}]{\includegraphics[scale=0.27]{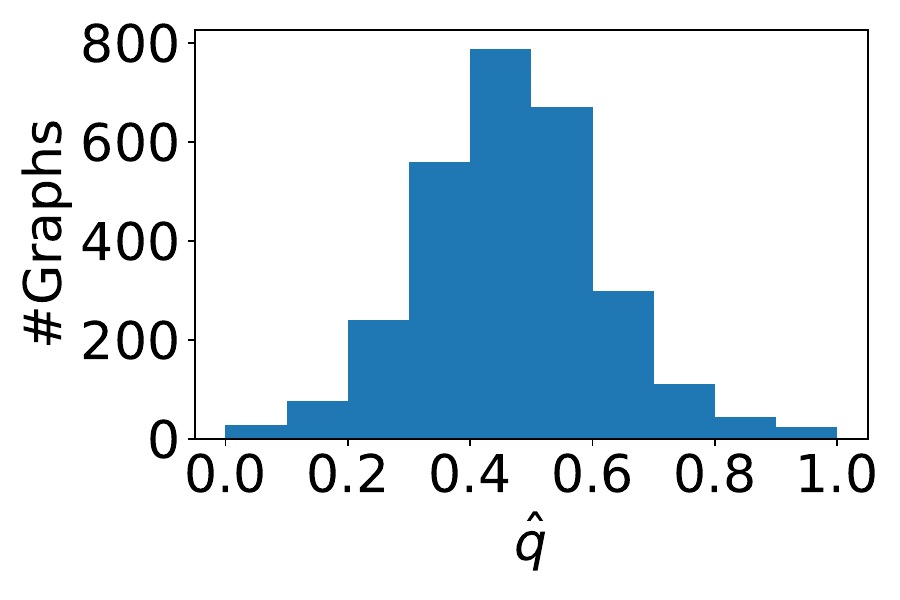}}
    \caption{(a): the number of Kialo reply trees as a function of their size (number of nodes). 
    (b): a histogram counting the number of Kialo reply trees with a given fraction of support $\hat{q}$.
    }
    \label{fig:hist_q}
\vspace{-0.5cm}
\end{figure*}

Figure~\ref{fig:hist_q} provides some basic statistics about the Kialo data. Figure~\ref{subfig-1-hist_q} shows the distribution of the sizes of the reply trees. On each topic or subdiscussion, there is a reasonable amount of debate, with a mean (median) of  204 (68) arguments (standard deviation 463).  
Figure~\ref{subfig-2-hist_q} calculates the fraction $\widehat{q}$ of replies that are supportive as opposed to attacking their parent argument. It appears that Kialo debates are typically balanced, with the vast majority of reply trees having $0.4 \le q \le 0.6$.  Section \ref{sec:scale_free_graphs} builds on these, in identifying the locations of justified arguments.

\section{A Probabilistic Model of Justified Arguments in Synthetic Bipolar Argumentation Frameworks}\label{sec:probabilistic_model}
In this section we will analyse and model probabilistically the distribution of the location of justified arguments in BAFs based on various kinds of reply trees. Our main aim will be to characterise the probability that we will find justified arguments at a given level of a reply tree (Section \ref{sec:prob_model_intro}). We will first study reply trees with homogeneous in-degree distributions (Section \ref{section:homogeneous}) and obtain an analytical model that allows us to understand how the levels of support or attack will affect the distribution of the locations of the justified argument. We will then study this in non-homogeneous in-degree trees, which better approximate the discussions we find in Kialo (Section \ref{sec:scale_free_graphs}).

\subsection{The Probability of Being Justified Given the Level}\label{sec:prob_model_intro}

Recall that each discussion is a directed tree, with the original post as the root, and a directed edge from each reply to the node being replied to. A reply can attack or support its parent, and we represent this by a negative or positive edge (see Equation \ref{eq:adj_mat_pos_neg} below). Let $q\in\sqbra{0,1}$ denote the probability that a reply is supporting and $1-q$ the probability that a reply is attacking.
Let $N\in\nat$ denote the depth of a given reply tree, i.e. the length of the longest path from root to leaf. 
For a given reply tree, let $0\leq h\leq N$ be an integer denoting an arbitrary level in the reply tree and $p(k|h)$ be the probability of a node at level $h$ having $k$ child nodes at level $h+1$ that reply to it. Apart from leaf nodes, which by definition cannot have any children and therefore no replies, all other nodes can have an arbitrary number of replies.
We wish to calculate the probability $p_h\in\sqbra{0,1}$, that a node at level $h$ will be a justified argument.   Recall from Section \ref{sec:background} that  unrebutted arguments are justified by default, so  the leaves in the reply tree are justified arguments. Given that the level $N$ is populated only by leaves. $p_N=1$. For internal nodes (i.e. nodes with $h < N$) to be justified, all of their child nodes (at level $h+1$) that support them must be justified, and all the child nodes that attack must be unjustified arguments or they should be leaves. Therefore, the expression for $p_h$ is
\begin{equation}\label{eq:prob_level_winning}
p_h=\sum_{k=0}^\infty\sqbra{qp_{h+1}+(1-q)(1-p_{h+1})}^kp(k|h).
\end{equation}
Given a set of arguments $A$ and  a set of justified arguments $G \subseteq A$, define a \textit{state function} $s:A\to\set{\pm 1}$ where $s_i$ is such that $s_i=1$ means $i\in G$, i.e. argument $i$ is justified, while $s_i=-1$ means $i\notin G$, i.e. argument $i$ is unjustified. The value of $s_i$ will be assigned iteratively to all $i\in A$ starting from level $N$ via the following rule:

\begin{equation}\label{winning}
s_i=
\begin{cases}
1 & \pair{\forall j\in A}J_{ij}=0\\
\min_{j\in A} J_{ij}s_j & \text{else}\\
\end{cases},
\end{equation}
where $\pair{J_{ij}}_{i,j\in A}$ is a matrix of size $\abs{A}\times\abs{A}$ defined as

\begin{align}\label{eq:adj_mat_pos_neg}
J_{ij}:=
\begin{cases}
-1 & (i,j)\in R_{att}\\
1 & (i,j)\in R_{sup}\\
0 & \text{else}.
\end{cases}.
\end{align}
Intuitively, the first case of Equation  \ref{winning} is the case of a leaf node which is justified by default. The second case assigns $s_i=1$ if  all reply nodes $j$ are either supporting reply nodes (i.e. $(i,j) \in R_{sup}$) and also justified nodes themselves (i.e. $s_j = 1$), or they are attacking nodes but are unjustified, and therefore their attack is invalid (i.e. $(i,j) \in R_{att}$ and $s_j = 1$). If neither of these conditions hold, there is at least one supporting node that is unjustified or one attacking node that is justified, which in turn means that node $i$ is not justified  and $s_i=-1$. 
We can now calculate the frequency of justified arguments at level $h$. This quantity, averaged over an ensemble of reply trees with the same degree distribution and in the same class of support $q$, will be our estimator of the probability $p_h$:
\begin{equation}
\widetilde{p}_h:=\frac{1}{2}\ang{\frac{\sum_{i_h\in A}(s_i+1)}{\sum_{i_h\in A}|s_i|}},
\end{equation}
where $|\cdot|$ indicates the absolute value, $i_h$ are the nodes in level $h$, $\langle \cdot \rangle$ is the average over the ensemble of trees and $\sum_{i_h\in A}|s_i|$ is the number of arguments at level $h$. In the next two subsections we will focus our analysis of the distribution of justified arguments on two kinds of graphs: trees with homogeneous in-degree distribution (Section \ref{section:homogeneous}) and scale-free trees (Section \ref{sec:scale_free_graphs}).

\subsection{Reply trees with homogeneous in-degree}\label{section:homogeneous}
A digraph with \textit{homogeneous in-degree distribution} is one where the degree distribution is the same for all the nodes. In the context of reply trees, this means that the distribution of the numbers of children (replies) does not vary across different levels (except for the deeper level, where the are no children): i.e. $\forall 0\leq h < N\ p(k|h)=p(k).$ 

As mentioned before, leaf nodes are unrebutted arguments and therefore considered to be justified in  bipolar argumentation frameworks. The theorem below obtains the probabilities $p_h$ of an argument among internal (non-leaf) nodes being justified at level $h$:

\begin{theorem}\label{thm:three_behaviours}
Let $p_h$ be the probability of being justified of an internal node at level $h < N$, given by Equation ~\ref{eq:prob_level_winning}. 
\begin{enumerate}
    \item If $q=\frac{1}{2}$ then 
    \begin{equation}
                    p_h=p_{h+1} \ \ \ \ \ \ \ \mbox{for}\ h \in [0,N-1]
    \end{equation} 
    \item If $q<\frac{1}{2}$ then 
    \begin{eqnarray}
p_{N-2m}&>& p_{N-2m+1} \ \ \ \ \ \ \ \mbox{for}\ m \in [0,N/2]  \\
p_{N-2m-1}&<& p_{N-2m} \ \ \ \ \ \ \ \ \, \mbox{for}\ m \in [0,(N-1)/2] 
\end{eqnarray}
\item If $q>\frac{1}{2}$ then 
\begin{equation}
p_{h} <  p_{h+1} \ \ \ \ \ \ \ \mbox{for}\ h \in [0,N-1]
\end{equation}
\end{enumerate}
\end{theorem}

\begin{proof}
We prove each case in turn.
\begin{enumerate}
\item If $q=\frac{1}{2}$ then $qp_{h+1}+(1-q)(1-p_{h+1})=\frac{1}{2}$. Therefore
\begin{align}\label{eq:p_h_indep_of_h}
p_h=\sum_{k=0}^\infty\frac{p(k)}{2^k}=p_{h+1} \ \ \ \ \ \ \ \mbox{for}\  h \in [0,N-1]
\end{align}
So irrespective of the in-degree distribution $p(k)$, $p_h$ will not depend on $h$.

\item If $q<\frac{1}{2}$, then
    
\begin{eqnarray}
p_{h}&>& p_{h+1} \ \ \ \ \ \ \ \mbox{if}\  p_{h+2}> p_{h+1}\text{ and} \label{2} \\
p_{h}&<& p_{h+1} \ \ \ \ \ \ \ \mbox{if}\  p_{h+2}< p_{h+1}.\label{3}
\end{eqnarray} 
    
\noindent If $p_{h+2}>p_{h+1}$, then
\begin{eqnarray}
\label{q<0.5_1}
p_{h}&=& \sum_{k=0}^\infty [qp_{h+1}+(1-q)(1-p_{h+1})]^kp(k) \\
&>& \sum_{k=0}^\infty [qp_{h+2}+(1-q)(1-p_{h+2})]^kp(k) = p_{h+1}.
\end{eqnarray}
This is because
\begin{eqnarray}
&\ & qp_{h+1}+(1-q)(1-p_{h+1})> \nonumber\\
&\ &qp_{h+2}+(1-q)(1-p_{h+2}),  \label{q<0.5_2}\\
&\ &-q(1-p_{h+1})+(1-q)(1-p_{h+1})>\nonumber\\
&\ &-q(1-p_{h+2})+(1-q)(1-p_{h+2})\label{q<0.5_3}\\
&\ & (1-2q)(p_{h+2}-p_{h+1})>0, \label{q<0.5_4}
\end{eqnarray}
since $q<\frac{1}{2}$ and $p_{h+2}>p_{h+1}$.

If instead $p_{h+2}<p_{h+1}$, then reasoning identically to the above implies that $p_{h}<p_{h+1}$. From our initial condition $p_N=1$ and
$$p_{N-1}=\sum_{k=0}^{\infty}q^kp(k)<p_N=1$$ we obtain an oscillating trend of $p_h$ as a function of $h$: $p_{N-2}>p_{N-1}$ as $p_{N-1}<p_N$, $p_{N-3}<p_{N-2}$ as $p_{N-2}>p_{N-1}$, and so on.

\item Finally, if $q>\frac{1}{2}$, then for all $0\leq h\leq N-1$,
\begin{equation}\label{1}
p_{h} <  p_{h+1},
\end{equation}
because $(1-2q)<0$, therefore the opposite of Equation  \ref{q<0.5_4} holds: if $p_{h+1}<p_{h+2}$, which holds true for $p_{N-1}<p_N=1$, then $p_{h}<p_{h+1}$, and so on monotonically.
\end{enumerate}
This shows the result.
\end{proof}

\begin{figure*}
\centering
\subfloat[\label{subfig-1-Poisson}]{\includegraphics[scale=0.27]{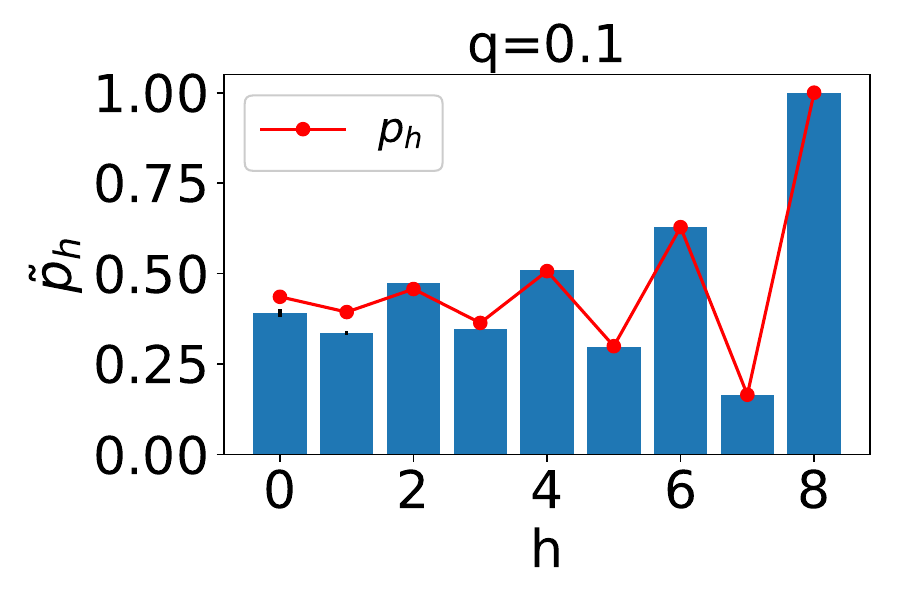}}
\subfloat[\label{subfig-2-Poisson}]{\includegraphics[scale=0.27]{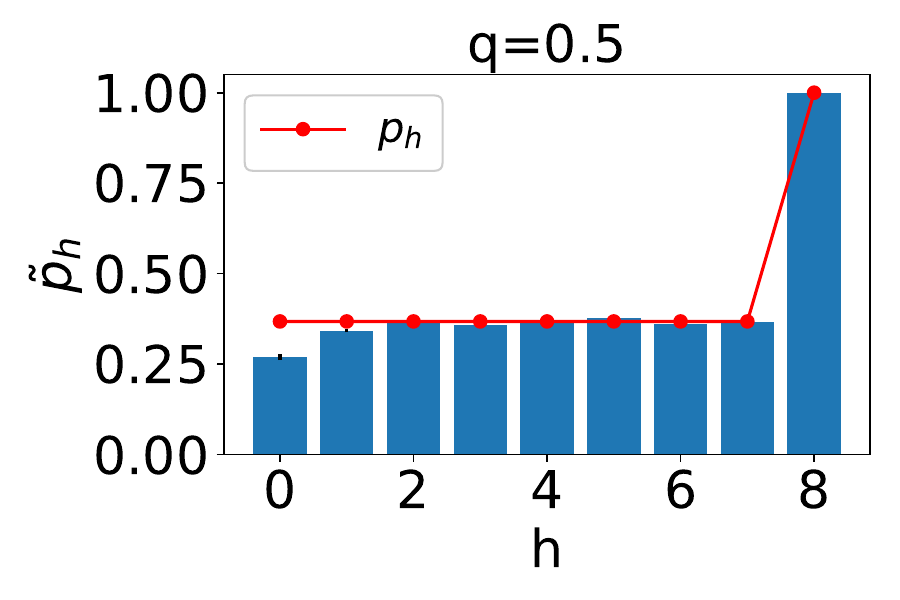}}
\subfloat[\label{subfig-3-Poisson}]{\includegraphics[scale=0.27]{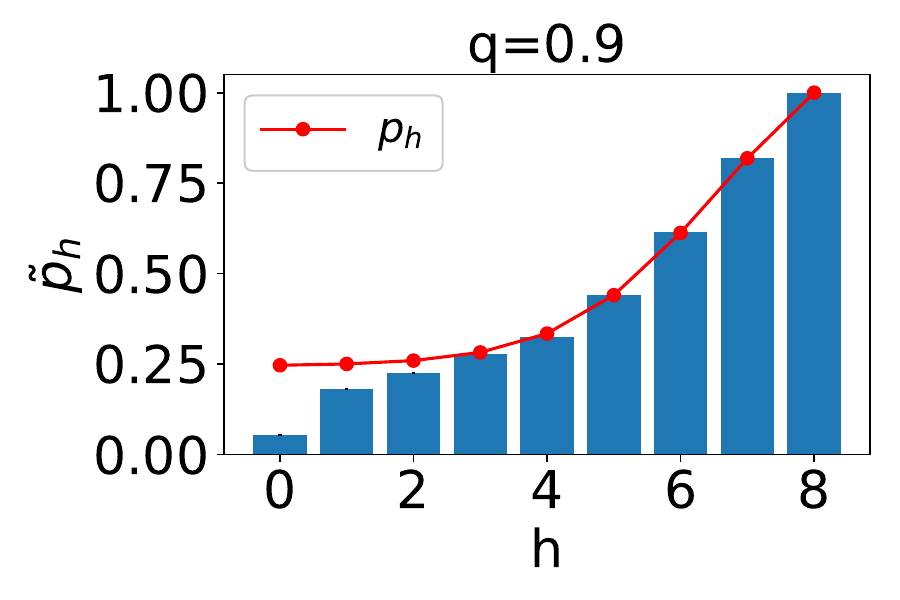}}
\caption{The figures show different values of $\widetilde{p}_h$ when the support probability $q$ is changed (here, $q\in\set{0.1,0.5,0.9}$). The degree distribution of the trees is Poisson with rate $\lambda=2$ for all levels. The red dots represent the theoretical prediction of the probability of an argument being justified at a certain level from Theorem~\ref{thm:three_behaviours}.}
\label{fig:Poisson}
\end{figure*} 

Intuitively, the theorem suggests there are three classes of behaviours for different values of $q$:
\begin{enumerate}
\item when $q<\frac{1}{2}$, there is a high probability that a reply is an attack. In this situation, the expected fraction of justified arguments at a given level is determined by the parity of the level, i.e. whether the path length from the leaf arguments or nodes (who are default justified) to a given argument (node) is even or odd. This is easiest to visualise in the extreme case when all arguments are attacking ($q=0$). This corresponds to the classical argumentation framework~\cite{Dung:95}. When there is a thread of replies (arguments) attacking each other, arguments at odd-length paths from the set of unattacked arguments $U$ are being \textit{indirectly attacked} by $U$, while even-length paths from the unattacked arguments are being \textit{indirectly defended} by $U$ (See \cite[Page 332]{Dung:95}).  This means that the proportion of justified arguments oscillates between levels -- i.e. the fraction of justified arguments increase and decrease from one level to the next.  
\item when $q=\frac{1}{2}$, and a reply is equally likely to be supportive or attacking, 
the expected fraction of justified arguments is the same for  all levels of the reply tree. This is because a node has the same probability of having a justified reply supporting or a unjustified reply attacking, independently on the level.
\item when $q>\frac{1}{2}$, the expected proportion of justified arguments is determined by how far a level stands from the deeper level. The farther a node is from this, the higher the probability that at least one of the child nodes in the subtree rooted at the node is justified and is attacking. Since most other edges in chains of replies will be supporting, this justified node will indirectly defeat all its ancestors. In other words, it becomes harder for nodes far away from the leaves to be a justified argument, and the probability of finding justified arguments increases monotonically as we go from away from the root.
\end{enumerate}

Figure~\ref{fig:Poisson} validates the results of Theorem~\ref{thm:three_behaviours} by showing these three behaviours in action in reply trees with a homogeneous (Poisson) in-degree distribution $p(k)=\frac{e^{-\lambda}\lambda^k}{k!}$, with $\lambda=2$.  Each of the three sub-figures in Figure \ref{fig:Poisson} is obtained by averaging the number of justified arguments of an ensemble of five hundred Poisson trees with depth $N=8$ and varying levels of support  $q\in\set{0.1,0.5,0.9}$. The theoretical estimates from Theorem~\ref{thm:three_behaviours} of the fraction of justified arguments at a given level appears to be in good agreement with what we observe from simulations. In particular we can recognise a transition at $q=\frac{1}{2}$ between the probability of an argument to be justified being driven by the parity of the distance from the deeper level (for $q<\frac{1}{2}$) and a probability of an argument to be justified that rises monotonically with the distance from the root (for $q>\frac{1}{2}$).

Theorem~\ref{thm:three_behaviours} has  implications both from a platform perspective and the users' perspective: When levels of support are high ($q \gg 0.5$),  the end of a conversation becomes much more important in determining the justified arguments of the discussion. So for readers, there is little point in following early comments, and the best strategy for platforms is to present comments in reverse chronological order, so that users see more of the justified comments first. In contrast, when there are high levels of attack, the justified arguments are more equally distributed across different levels, and users still benefit from reading early comments. Note that these insights apply mainly to evolving discussions where new comments are still being added. When a discussion thread has received all its comments, the rules of BAF can be used to clearly determine justified arguments.

\subsubsection{Oscillation Amplitude and the Decay of $p_h$}

Now that we have characterised the trend of the probability $p_h$, we would like to know how large are the oscillations of $p_h$ for $q<\frac{1}{2}$ and how steep is its decay when $q>\frac{1}{2}$. In particular, we will answer these questions in relation to the size of probability $p_{leaf}:=p(0)$; this is the probability for a node being a leaf. We will see that this analysis will be of particular interest in order to understand the distribution of justified arguments in Kialo data.
Consider a (homogeneous in-degree) tree with a certain support probability $q$. We can rewrite Equation \ref{eq:prob_level_winning} as follows:
\begin{eqnarray}
p_N &=& 1\\
p_{h}&=& p(0)+\sum_{k=1}^\infty [qp_{h+1}+(1-q)(1-p_{h+1})]^kp(k). \label{sum}
\end{eqnarray}
\noindent As $\sqbra{qp_{h+1}+(1-q)(1-p_{h+1})}^k\leq qp_{h+1}+(1-q)(1-p_{h+1})$ for all $k\geq 1$, we have that
\begin{eqnarray}
p_{h}&\le & p(0) + \sqbra{qp_{h+1}+(1-q)(1-p_{h+1})}(1-p(0))\\
&=:& p_h^{max}\label{up}\\
p_{h}&\ge &p(0) =: p_h^{min}\label{down}.
\end{eqnarray}

\begin{figure}
    \subfloat[\label{subfig-1}]{\includegraphics[scale=0.3]{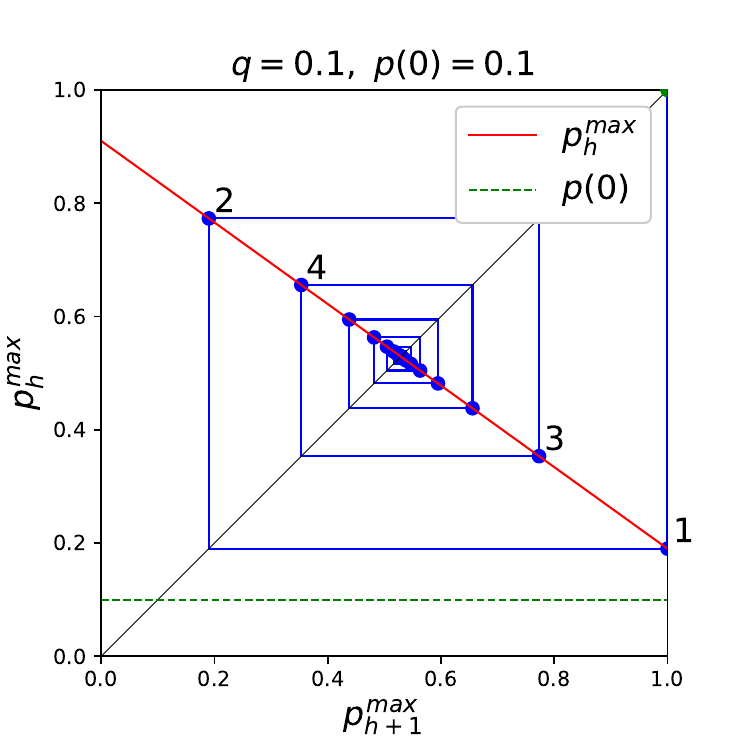}}
     \subfloat[\label{subfig-2}]{\includegraphics[scale=0.3]{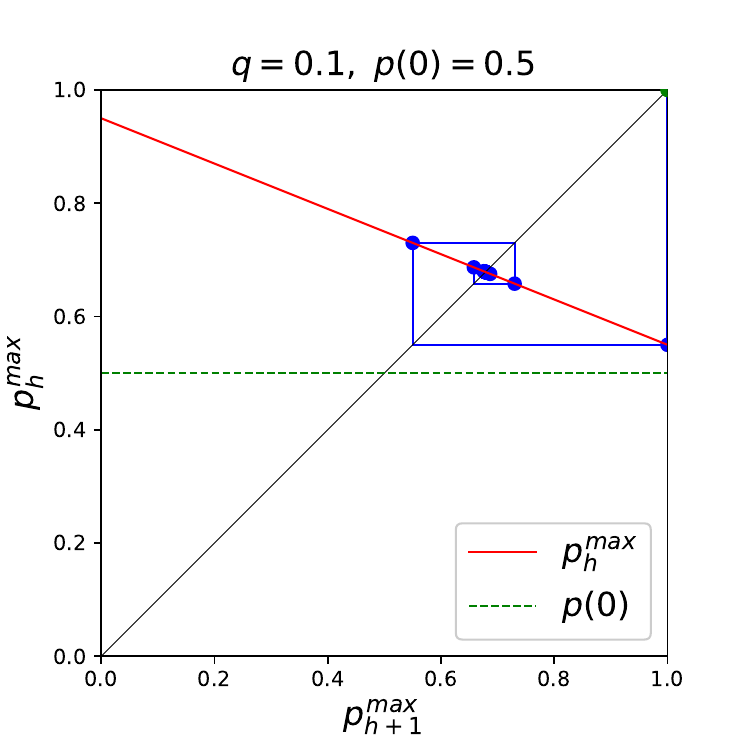}}
     
    \subfloat[\label{subfig-3}]{\includegraphics[scale=0.3]{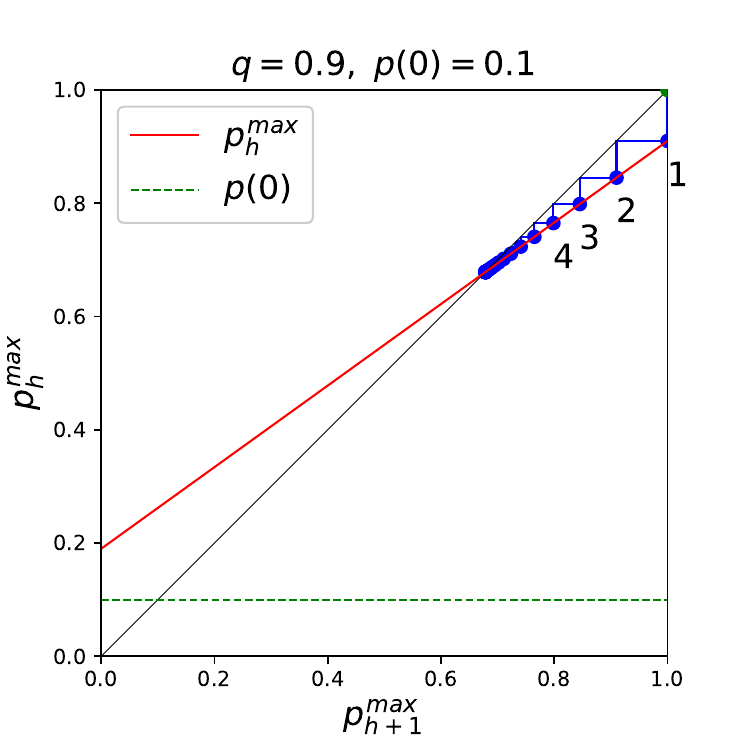}}
    \subfloat[\label{subfig-4}]{\includegraphics[scale=0.3]{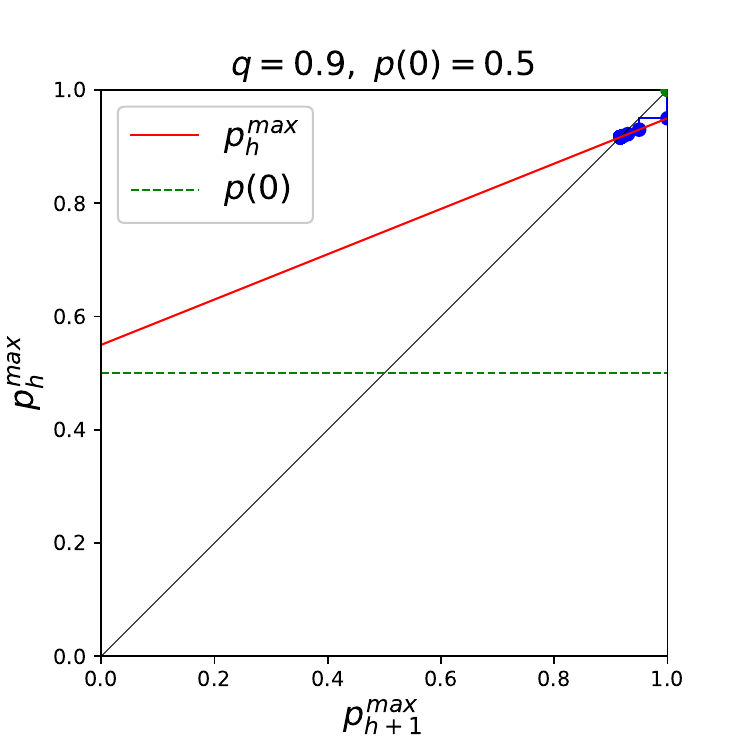}}
    \caption{Cob-webbing solution of the upper bounds (blue dots) of the function $p_h^{max}$  (Equation \ref{p_hmax}) for $q<\frac{1}{2}$ and $p(0)=0.1$ (a), for $q<\frac{1}{2}$ and $p(0)=0.5$ (b), for $q>\frac{1}{2}$ and $p(0)=0.1$ (c) and for $q>\frac{1}{2}$ and $p(0)=0.5$ (d). We numbered the blue dots representing the first iterative solutions only for $p(0)=0.1$.}
    \label{fig:cobweb_1}
\end{figure}

Equations \ref{up} and \ref{down} provide an upper and a lower bound of the function $p_h$. The upper bound in Equation  \ref{up} is composed of two terms: The first one,  $p(0)$,  is the probability of a node to be a leaf and the other term  depends on the  probability $p_{h+1}$ of the nodes at the following level being justified arguments. This second term is responsible for the oscillations of the upper bound in function of the level for $q<\frac{1}{2}$, and the decrease of the probability of an argument being justified as function of the level for $q>\frac{1}{2}$ (similar to what we have seen in Equations \ref{2}, \ref{3} and \ref{1}). However the higher is the lower bound the smaller will be the amplitude of the oscillations and the decrease per-level, as $p_h$ will be squeezed between a large $p_h^{min}$ and 1. 
 This is shown in Figure \ref{fig:cobweb_1}. Figures \ref{subfig-1} and \ref{subfig-3} show systems with a relatively small $p(0)=0.1$, indicated by the green dashed line. Figures \ref{subfig-2} and \ref{subfig-4}  show systems with a relatively large $p(0)=0.5$. This will respectively determine large and small oscillations of $p_h$ in Figure \ref{subfig-1} and \ref{subfig-2}  and long and short decrease of $p_h$ in Figures \ref{subfig-3} and \ref{subfig-4}.
 The blue dots represent the iterative solution of the equation:
\begin{equation}\label{p_hmax}
p_h^{max}=p(0) + \sqbra{qp_{h+1}^{max}+(1-q)(1-p_{h+1}^{max})}(1-p(0)),
\end{equation}
which was obtained by assuming that the inequality in Equation \ref{up} is saturated, i.e. $p_h=p_h^{max}$. The red lines in Figure~\ref{fig:cobweb_1} represents the Equation \ref{p_hmax}. The iterative solutions of Equation \ref{p_hmax} indicated by the blue dots is obtained by projecting the points on this line to the diagonal black line. For example starting from $p_N=1$, we obtain the blue point 1 in Figure \ref{subfig-1}, which is a solution of Equation \ref{p_hmax} with initial $p_{h+1}=1$. the projection of this point on the diagonal is the starting point of the new iteration. This new starting point, corresponding to a new value of $p_{h+1}$, leads to the new solution of Equation \ref{p_hmax} and is indicated by the blue point 2. Note that the new $p_{h+1}$ is much smaller than the initial one, and also much smaller than the next value in the iteration represented by the blue point 3. As shown in Figure \ref{subfig-1} and \ref{subfig-2}, the oscillations thus produced are larger when the value of $p(0)$ is small. This representation of iterative solutions is called \textit{cob-webbing} \cite{muller2015methods}. 
The same cob-webbing procedure for $q>\frac{1}{2}$ is shown in Figure \ref{subfig-3} and \ref{subfig-4}. In this case we do not have oscillations but similarly we can see that when $p(0)$ is large there is a less pronounced decrease of $p_h$.
As a consequence of this we can conclude that the amplitude of the oscillations and the decrease of the solution of $p_h$ depend on the size of $p(0)$. This means that the number of unreplied comments in the reply tree (which determines $p(0)$) has a large impact on the behaviour of the probability of an argument being justified. We will apply this analysis and result when analysing Kialo discussions in the next section.

\subsection{Non-Homogeneous Reply Trees}\label{sec:scale_free_graphs}
Here, we consider trees that have in-degree probability distributions which are non-homogeneous across different levels; we will call these \textit{non-homogeneous} reply trees. These trees more closely approximate the empirical data from Kialo but given the in-degree distribution is not the same for all the levels, we are not able to provide a closed-form solution for the probability of an argument being justified as a function of the level and in-degree. Instead, we study the distribution of justified arguments by generating an ensemble of scale-free trees (an example of non-homogeneous trees ~\cite{katona2006levels} often appearing in social processes), and examine the patterns found in comparison to Kialo discussions.

A common way to generate scale-free graphs is using preferential attachment \cite{barabasi1999emergence}.  To generate scale-free \emph{trees} with preferential attachment, we follow the method of Krapivsky and Ridner~\cite{PhysRevE.63.066123}:  
At each step, we add a new node and connect it to an existing node $i$ with probability $\pi_i=\frac{w_i}{\sum_j w_j},$
where $w_i$ is the degree  
of the node $i$ and the sum at the denominator runs over all the existing nodes. Intuitively, each new node is attached  with higher probability to a node with high degree, leading to preferential attachment.

For each simulation we generate 1000 random trees of size 100. Each edge is assigned at random  to be a support with a probability $q$ and with probability $1-q$ to be an attack.
 The average of the observables has been done between levels with the same distance from the deepest level of their tree.
This is because we expect that the nodes at the same distance from the deepest level of their tree have a comparable in-degree distribution. In all the figures that we will show in this section and the next one, the number of levels on the horizontal axis will correspond to the depth of the highest tree analysed. Given the scale-free nature of our graphs, there will be  a number of short trees with many leaves and few long trees, leading to fewer trees to average on, at relatively higher levels ($h>10$). 
To maintain statistical significance, we only report probabilities of arguments being justified for levels which have at least ten trees with nodes at that level. 

\begin{figure*}
    \centering
    \subfloat[\label{subfig-1-scalefree}]{\includegraphics[scale=0.27]{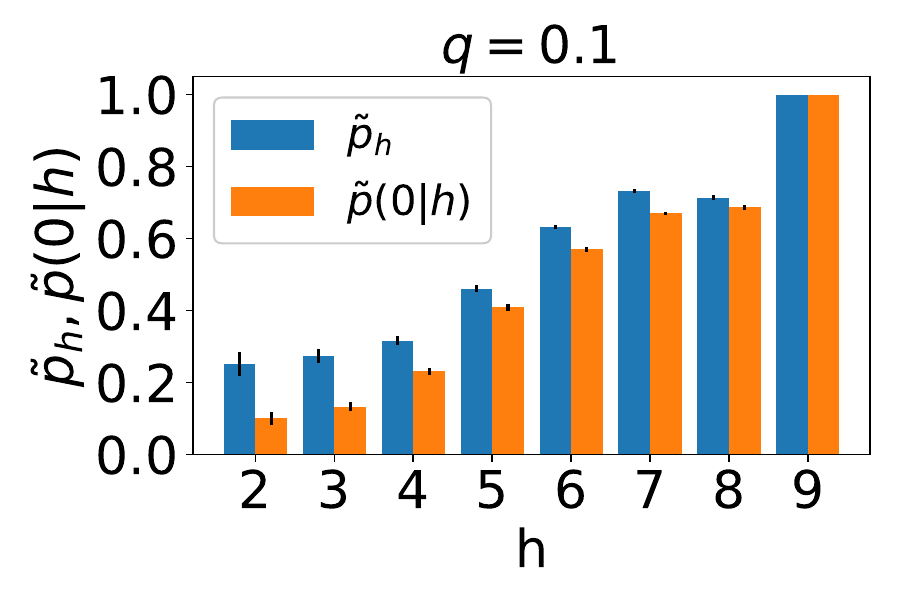}}
    \subfloat[\label{subfig-2-scalefree}]{\includegraphics[scale=0.27]{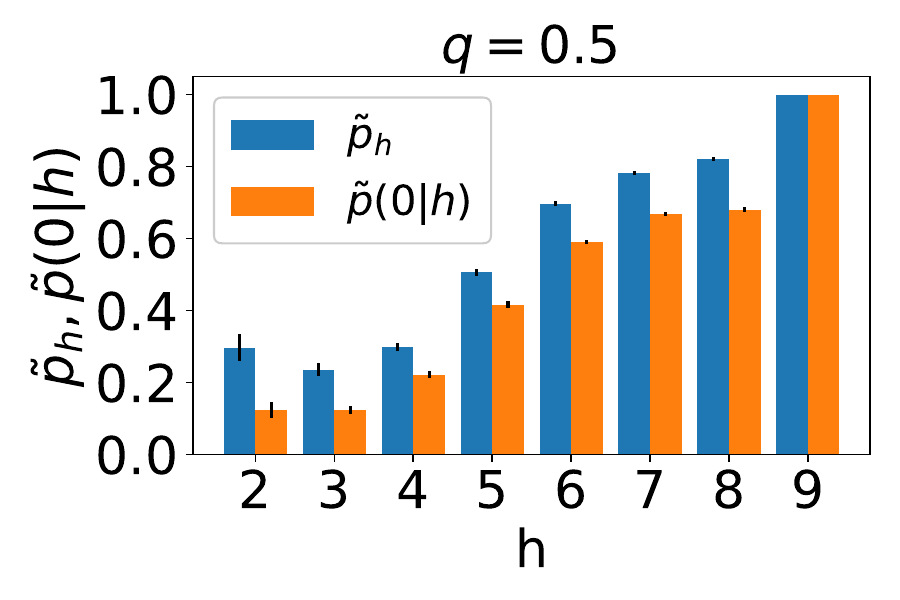}}
    \subfloat[\label{subfig-3-scalefree}]{\includegraphics[scale=0.27]{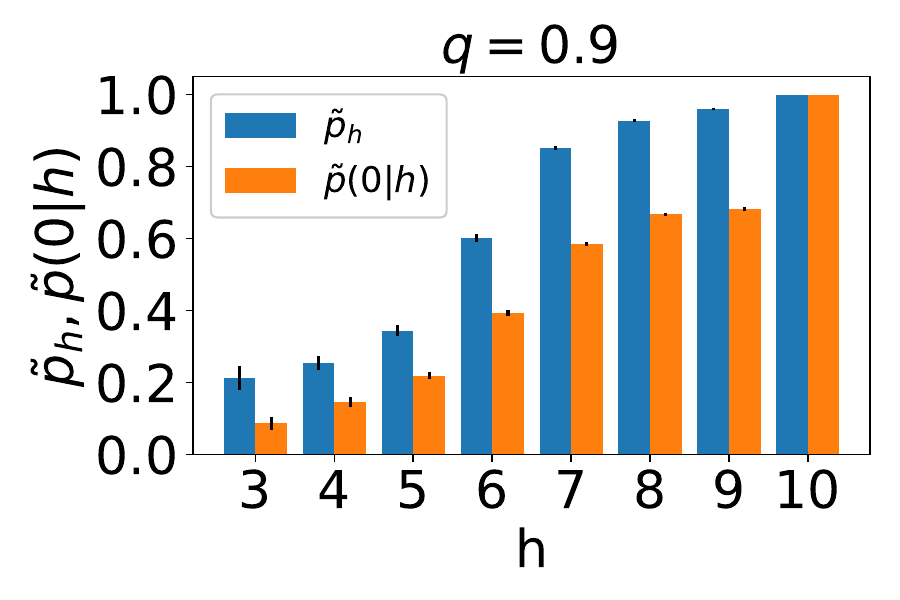}}
    \caption{Estimated probability of an argument being justified per level in scale-free synthetic graphs for different levels of support in the graph, compared to $\widetilde{p}(0|h)$, the estimated probability of having leaves at level $h$.}
    \label{fig:scalefree}
    \vspace{-0.5cm}
\end{figure*}
\begin{figure*}
    \centering
    \subfloat[\label{subfig-1-stat}]{\includegraphics[scale=0.27]{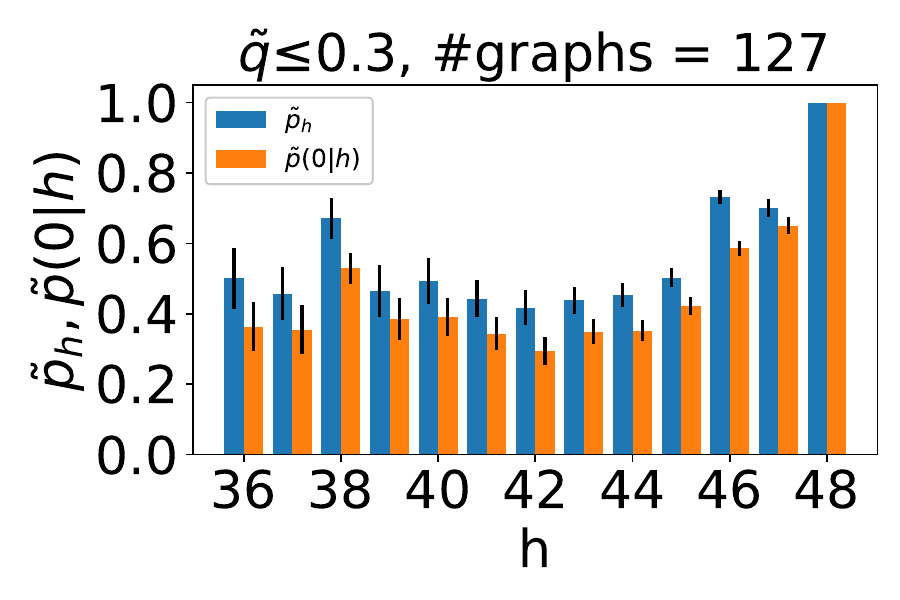}}
    \subfloat[\label{subfig-2-stat}]{\includegraphics[scale=0.27]{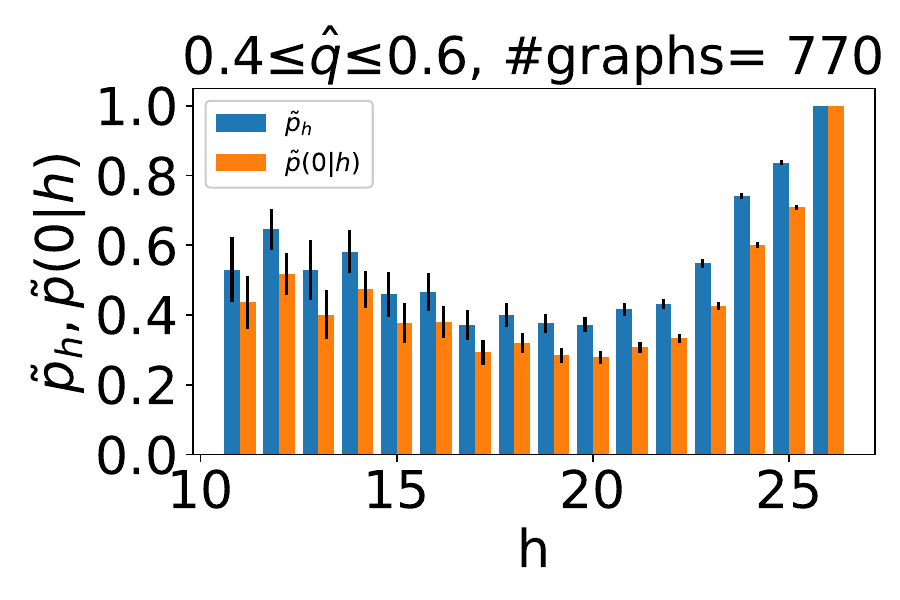}}
   \subfloat[\label{subfig-3-stat}]{\includegraphics[scale=0.27]{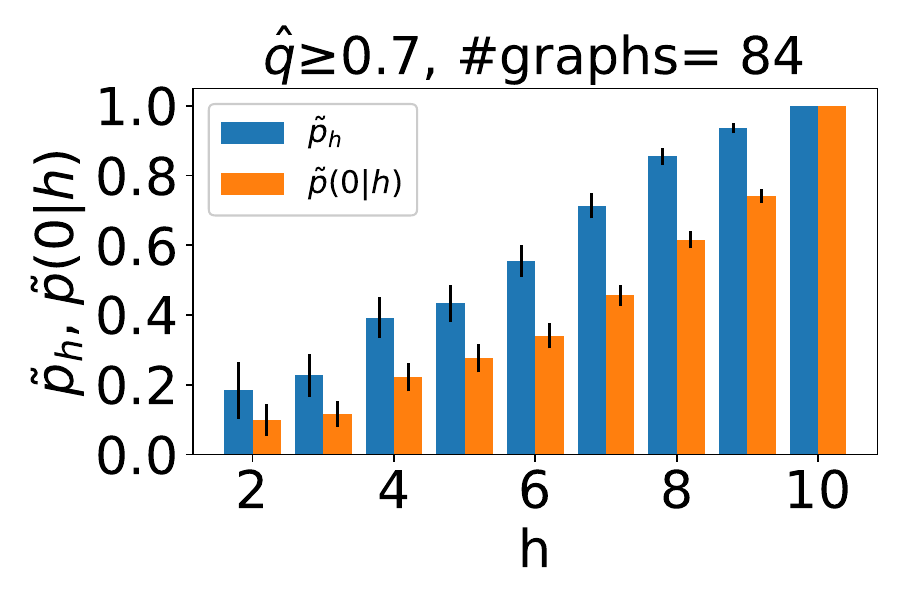}}
    \caption{Estimated probability of an argument being justified per level in Kialo discussions compared to $\widetilde{p}(0|h)$, the estimated probability of arguments being leaves at level h, for different levels of support.}
    \label{fig:stat}
    \vspace{-0.5cm}
\end{figure*}

In Figure \ref{fig:scalefree}, we plot (in blue) how the probability $\widetilde{p}_h$ varies with the level $h$. We observe that unlike the homogenous degree distribution case, the justified arguments are overwhelmingly found at higher levels, away from the root. This happens regardless of the level of attack or support (i.e. regardless of the value of $q$). We next plot (in orange) the distribution of default justified arguments, i.e. leaf nodes. Since leaf nodes have $k=0$ children, this distribution is computed as $\widetilde{p}(0|h)$. For all values of $q$, we can see that the distribution of justified arguments at level $h$,  $\widetilde{p}_h$, closely follows the distribution of leaves $\widetilde{p}(0|h)$. In other words, \textit{the distribution of justified arguments is dominated by the large numbers of default justified arguments or leaves}. 

Next, we examine real-world discussions from Kialo in Figure~\ref{fig:stat}. As mentioned in Section \ref{sec:Kialo}, Kialo is evenly balanced between attacking and supporting comments, with a vast majority of discussions having $0.4 < \widehat{q} < 0.6$, and very few highly supportive or highly attacking conversations. For highly supportive graphs, with $0.8 < \widehat{q} < 1$, we see that the farther an argument is from the root, higher is its probability of being justified. For graphs with this amount of support we can observe this behaviour in both homogeneous and non-homogeneous graphs. For $0.4 <q<0.6$ and $q<0.3$ we instead cannot recognize the behaviours seen in the homogenous case. We can observe instead that whatever is the level of support in the graphs, as in scale-free trees, the distribution of justified arguments at a given level $p_h$, closely follow the distribution of leaves in the graph $p(0|h)$.

\section{Removing leaves from the count of justified arguments}\label{sec:removing_leaves}

Our study of synthetic reply networks and comparison with Kialo data (Section \ref{sec:probabilistic_model}) seems to highlight that comments that have the last word (i.e. the leaf comments in discussion trees) represent a determining factor in establishing the rest of the justified arguments. This is consistent with argumentation theory, which assumes that arguments that have the last word are justified by default (Section \ref{sec:background}). 

However, it is not fully clear whether this is appropriate for online discussions. Although one may argue that comments which are spurious or false are rarely left unchallenged in vigorous online debates, and therefore the leaf arguments can be treated as justified arguments, it can also be argued that comments that have been answered, and so have been evaluated positively or negatively by others, are more representative of the truth and should have a greater importance than those that have not yet been replied to. 

In this section, we consider a conservative approach, where we use the machinery of argumentation theory to identify the justified arguments, but do not include the leaf nodes of a reply tree in the \textit{count} of justified arguments. In other words, we only consider those arguments that have had a chance to be supported or attacked by at least one other argument.
Given a reply tree with $n_h$ nodes at level $h$ and a distribution of leaves given by 
\begin{eqnarray}
\widehat{p}(0|h)&:=& \frac{\# \mbox{leaves at level\ } h}{n_h},\label{single_tree2}
\end{eqnarray}
we previously defined the probability of an argument being justified at that level as: 
\begin{eqnarray}
\widehat{p}_h&:=& \frac{\# \mbox{justified arguments at level\ } h}{n_h}\label{single_tree1}
\end{eqnarray}

In this section, we will instead compute the probability $\widetilde{p}_h^{nl}$ of \textit{non-leaf} justified arguments at a level $h$ by removing the count of the leaves ($n_h p(0|h)$) from the count of justified arguments at that level ($n_h \widehat{p_h}$):
\begin{eqnarray}
\widehat{p}^{nl}_h&=& \frac{\# \mbox{justified arguments at level\ } h-\# \mbox{leaves at level\ } h}{n_h-\# \mbox{leaves at level\ } h} \nonumber\\ \nonumber
&=&\frac{\widehat{p}_h-\widehat{p}(0|h)}{1-\widehat{p}(0|h)}.
\end{eqnarray}
With this change, the estimated probability of arguments being justified per level is

\begin{equation}
    \widetilde{p}_h^{nl}=\left \langle \frac{\widehat{p}_h-\widehat{p}(0|h)}{1-\widehat{p}(0|h)} \right \rangle, 
    \label{pnlHAT}
\end{equation}
where the average $\langle \cdot \rangle$ is over an ensemble of graphs with the same degree probability and the same level of support $q$. We will examine how this new definition affects the distribution of justified arguments:

\paragraph{Homogeneous in-degree distributions} 
For reply trees with homogeneous in-degree distributions, the distribution of leaves does not change with level (for all $0\leq h< N$, $p(0|h)=p(0)$), so the shape of the probability of the non-leaf arguments being justified $p_h^{nl}$ as a function of the level would not differ much from the old probability distribution $p_h$: 
\begin{equation}
    \widetilde{p}_h^{nl}=\left \langle \frac{\widehat{p}_h-\widehat{p}(0)}{1-\widehat{p}(0)} \right \rangle. 
\end{equation}

\paragraph{Non-homogeneous in-degree distributions} In general (e.g. in Kialo or scale-free trees),  the estimated probability of leaves per level has a non-trivial dependence on the level $h$, and therefore  $\widetilde{p}_h^{nl}$ behaves differently from $\widetilde{p}_h$. 
We can compute $\widetilde{p}_h^{nl}$ by separating the contribution of leaves from that of the other comments in Equation (\ref{eq:prob_level_winning}):
\begin{eqnarray}\label{noleaves}
    p_h^{nl}&=&\sum_{k=1}^\infty \left[q\left(p(0|h+1)+(1-p(0|h+1))p_{h+1}^{nl}\right) \right.\nonumber\\
    &\ &+ (1-q)(1-p(0|h+1))\left.\left(1-p_{h+1}^{nl}\right)\right]^kp(k|h),
\end{eqnarray}
where the first term is the probability of being supported by a leaf, or being supported by a non-leaf that is justified. The second term is the probability of being attacked by a non-leaf that is unjustified. Note that this time the sum over all replies to the node at level $h$ starts from $k=1$ in order to exclude the deeper level composed only by leaves (which would correspond to $k=0$).
Given that we do not have an analytical formula for $p(k|h)$, we will approximate the solution of Equation \ref{noleaves} using the fraction of replies per level of a single synthetic graph. We define this as
\begin{eqnarray}\label{eq:this_quantity}
\widehat{k}_h&=&\frac{\sum_{i_h} k_{i_h}}{n_h},
\end{eqnarray}
where $i_h$ indexes the nodes belonging to level $h$, and $k_{i_h}\in\nat$ is the in-degree of node $i_h$. Therefore, Equation \ref{eq:this_quantity} is the average in-degree of level $h$.

We can approximately estimate the new probability of an argument being justified per level substituting in Equation \ref{noleaves} $k$ with $\widehat{k}_h$ (Equation \ref{eq:this_quantity}) and $\widehat{p}(0|h)$ (Equation \ref{single_tree2}) to $p(0|h)$:
\begin{eqnarray}
p_h^{nl} &\approx& \Big \langle \left[q \left(\widehat{p}(0|h+1)+(1-\widehat{p}(0|h+1))p_{h+1}^{nl}\right) \right.  \nonumber\\
&\ & +  \left. (1-q)(1-\widehat{p}(0|h+1))\left(1-p_{h+1}^{nl} \right)\right]^{\widehat{k}_h} \Big \rangle \: . \label{noleaves_approx}
\end{eqnarray}

\begin{figure}
    \centering
    \subfloat[\label{subfig-1-no_leaves}]{\includegraphics[scale=0.27]{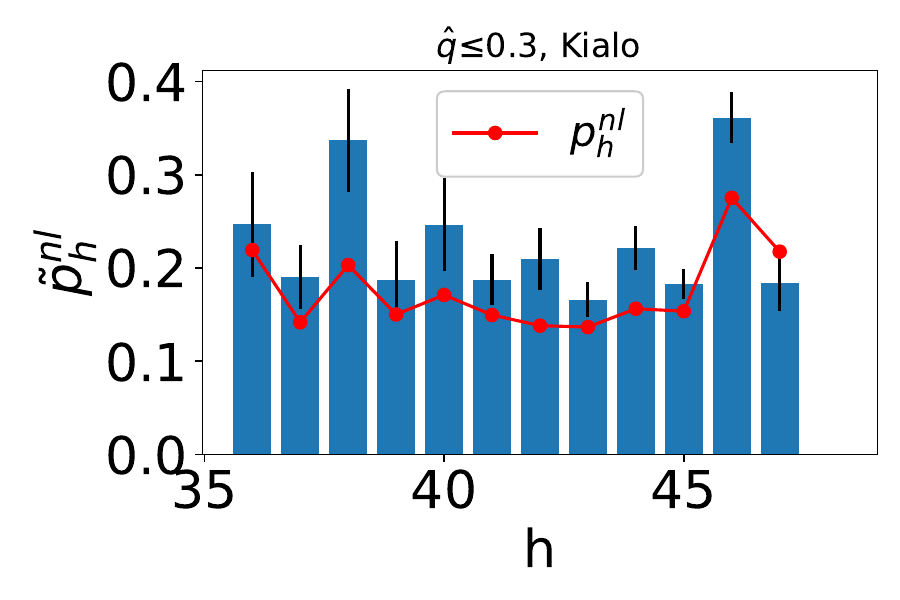}}
    \subfloat[\label{subfig-2-no_leaves}]{\includegraphics[scale=0.27]{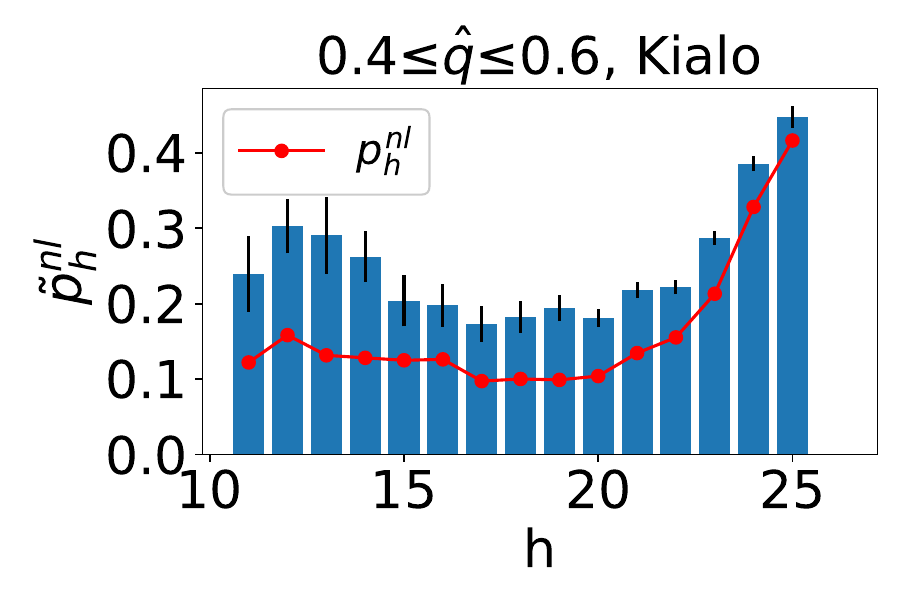}}
    \subfloat[\label{subfig-3-no_leaves}]{\includegraphics[scale=0.27]{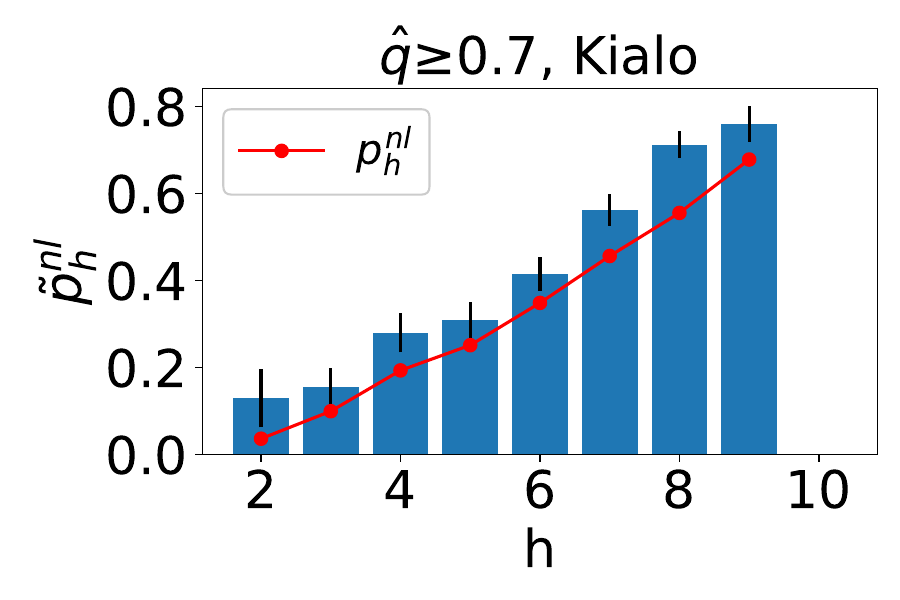}}
    
    \subfloat[\label{subfig-4-no_leaves}]{\includegraphics[scale=0.27]{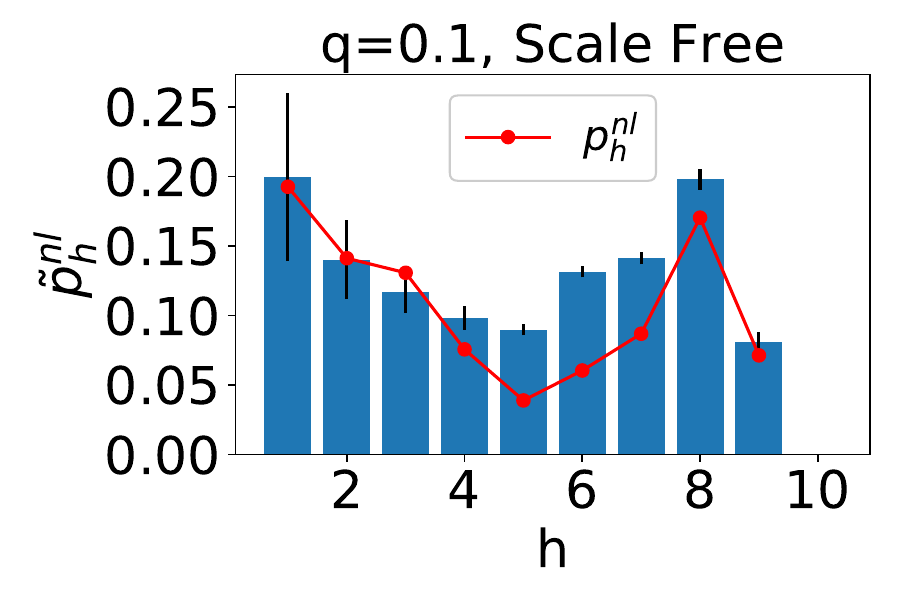}}
    \subfloat[\label{subfig-5-no_leaves}]{\includegraphics[scale=0.27]{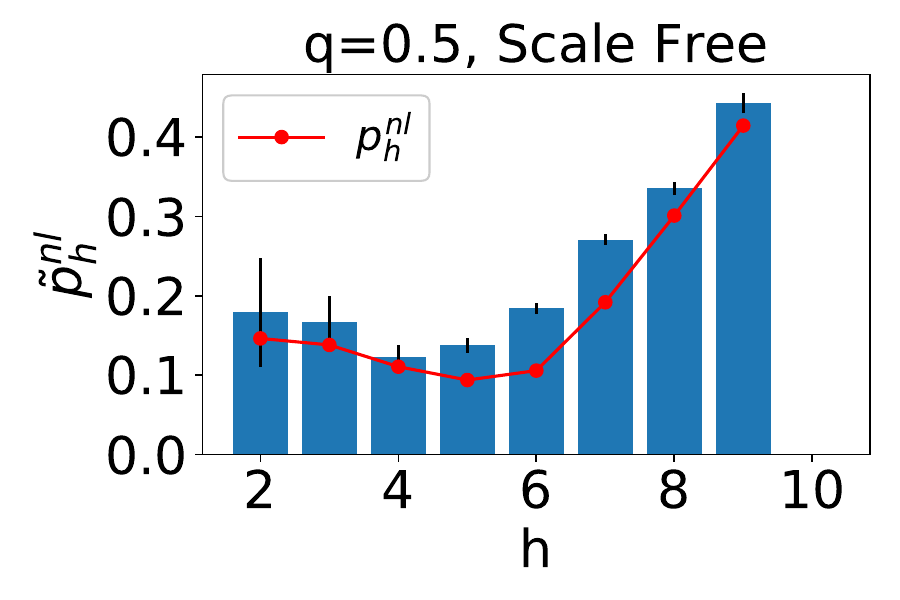}}
    \subfloat[\label{subfig-6-no_leaves}]{\includegraphics[scale=0.27]{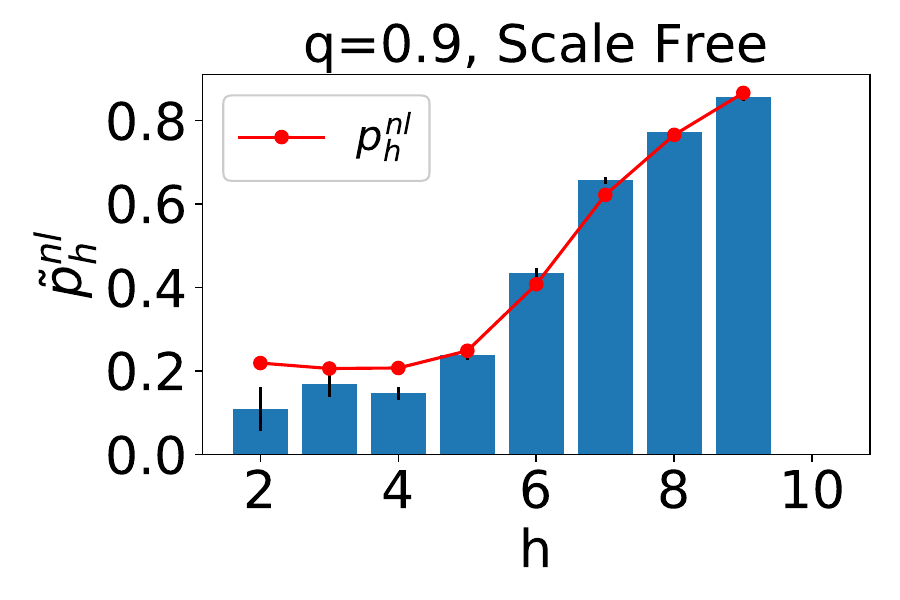}}
    \caption{Compare the estimated probability of non-leaf arguments being justified $\widetilde{p}_h^{nl}$ with its theoretical prediction (Equation~\ref{noleaves}) for Kialo discussions ((a), (b) and (c)) and scale-free synthetic graphs ((d), (e) and (f)) and different levels of support ($q$). The scale-free graphs have been generated as usual and the quantities are averaged over 1000 trees of size 50.}
    \vspace{-4mm}
    \label{fig:no_leaves}
\end{figure} 
\begin{figure}
    \centering
    \subfloat[\label{subfig-1-kh}]{\includegraphics[width=0.45\columnwidth]{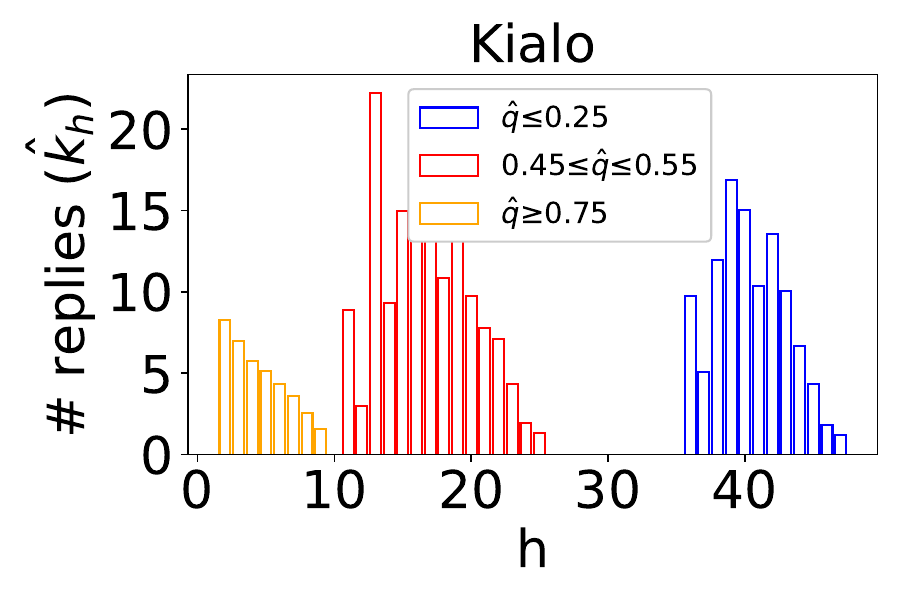}}
    \subfloat[\label{subfig-2-kh}]{\includegraphics[width=0.45\columnwidth]{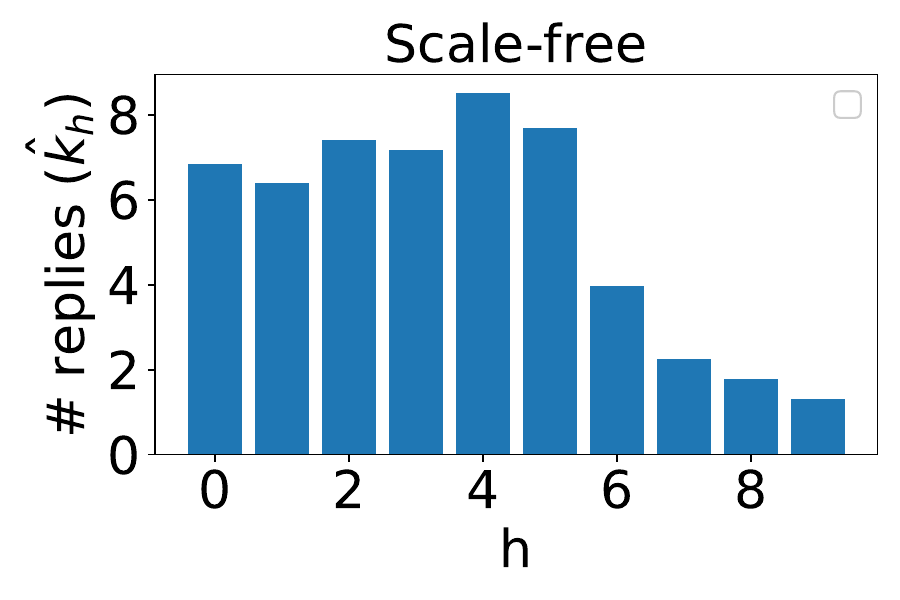}}
    \caption{Distribution of average number of replies per level ($\widetilde{k_h}$) for Kialo and Scale-free graphs}
    \label{fig:kh}
    \vspace{-0.5cm}
\end{figure}
To understand Equation~\ref{noleaves_approx}, we consider, as done in the previous section, three different regimes: perfectly balanced discussions ($q\approx \frac{1}{2}$); aggressive or acrimonious discussions ($q\approx 0$) and supportive discussions $q \approx 1$. 

\paragraph{Balanced discussions} 
For $q\approx \frac{1}{2}$, we can see that the formula simplifies and the probability of an argument being justified is given by:
\begin{equation}
p_h^{nl}\approx \ang{ q^{\widehat{k_h}}}. \label{q05_noleaves}
\end{equation}
In other words, the probability of non-leaf justified arguments depends solely on the number of replies an argument at a given level gets on average (i.e. on  $k_h$). Since $q<1$, more the number of replies, the \textit{lower} that $p_h^{nl}$ gets. We see this most clearly in scale-free trees: A general result about scale-free graphs \cite{katona2006levels} is that the levels with the highest number of nodes are expected to be in the middle of the graph when the number of levels is very large. In Figure $\ref{subfig-2-kh}$ we show $\widehat{k_h}$ averaged over 1000 scale-free trees of 100 nodes. Even for a short tree we can see that the middle level of the graph receive the highest attention. Thus, we expect to have a minimum of $p^{nl}_h$ in the middle of the tree, as observed in Figure \ref{subfig-5-no_leaves}.

In Kialo discussions (with $0.4 < \widehat{q} < 0.6$) we also have that the highest number of replies is found at the center of the graphs, but with a larger drop off as we get towards the root (See Figure \ref{subfig-1-kh}). In other words, deeper into a discussion thread, the arguments get lesser scrutiny and therefore fewer replies, leading to an increasing probability of arguments being justified as we go deeper into a discussion (as observed in Figure \ref{subfig-2-no_leaves}).  Whereas these are the normative justified arguments as defined by argumentation theory, there may be other human factors which drive this distribution of replies. For instance, user fatigue may lead to fewer replies deeper in a discussion thread (recall that balanced discussions have particularly large $h$ values in Kialo). Such factors need to be disentangled in future work. 

\paragraph{Supportive discussions}. 

In supportive discussions, when $q\gg 0$, we can identify three main ingredients from Equation \ref{noleaves_approx} in the determination of the justified arguments at a level $h$: the average number of replies $\widehat{k_h}$, the probability of a non leaf argument being justified $p^{nl}_{h+1}$ at level $h+1$ and the proportion of leaves per level $\widehat{p}(0|h)$. Using these three we are able to well predict the behaviour of the justified arguments, as we can see from Figure \ref{subfig-3-no_leaves}. 
It is important to notice that even if we removed the leaves in the calculation of justified arguments, their influence is an important ingredient in the determination of $p^{nl}_h$. Given the non-trivial dependence of $p^{nl}_h$ to the three ingredients discussed above, the system behaviour is not straightforward. However, we can notice from Figure \ref{fig:no_leaves} that the behaviour of Kialo graphs (Figure \ref{subfig-3-no_leaves}) does not change when the leaves are removed and it is the same as scale-free graphs (Figure \ref{subfig-6-no_leaves}) and homogeneous graphs (Figure \ref{subfig-3-Poisson}), with  $p_h^{nl}$ increasing as we go away from the root. 

\paragraph{Aggressive discussions}. 
In aggressive discussions $q \ll 1$ and we still have the same non-trivial dependence of $p^{nl}_{h}$ on $\widehat{k_h}$, $p^{nl}_{h+1}$ and $\widehat{p}(0|h)$. 
As before, Equation \ref{noleaves_approx} gives a good prediction of the behaviour of $p^{nl}$ and shows that leaves have an influence on the probability of non-leaf arguments being justified even if they are not counted in the set of justified arguments. 
In aggressive discussions, the behaviour of scale-free graphs (Figure~\ref{subfig-4-no_leaves}) appear to be totally different from the behaviour of Kialo graphs when $\widehat{q}<0.3$ (Figure~\ref{subfig-1-no_leaves}). 
 In fact, $\widetilde{p}_h^{nl}$ in Kialo discussions shows a clear oscillatory behaviour typical of homogeneous trees (Figure \ref{subfig-1-Poisson}).
The impossibility to see this oscillatory behaviour in $p_h$ (Figure \ref{subfig-1-stat}) can be explained by the presence of leaves in the count of justified arguments. In fact, looking at Equation \ref{p_hmax}, which was approximately describing the oscillation of the upper bound $p_{max}$ of $p_h$ in homogeneous trees with $p(0) \gg 0$, we can see that the oscillations amplitude was controlled by leaf probability $p(0)$ and dampened for large $p(0)$. If now we use the statistics of non-leaves justified comments formalized in Equation \ref{pnlHAT}, i.e subtract $p(0)$ from $p_{max}$, and we divide by $(1-p(0))$ (for $p(0) < 1$), we are left with:

\begin{align}
\frac{p_{max} - p(0)}{1-p(0)} = qp_{h+1}^{max} + \pair{1-q}\pair{1-p_{h+1}^{max}}.
\end{align}
The analysis of the distribution of non-leaf  arguments that are justified is in this case effectively zooming in on the level-by-level oscillations in the numbers of justified arguments that is expected in a homogeneous tree (see Figure \ref{subfig-1-Poisson}) rather than a scale-free non-homogeneous tree.

\section{Conclusions and Future Work}\label{sec:conc}

This paper applies ideas from bipolar argumentation theory and complex networks to an ensemble of synthetic reply trees where the nodes are arguments and the directed edges are attacking or supporting replies. We then built a model that calculates the probability that an argument will be justified in the debate given given its ``level'' or distance from the main thesis of the debate, i.e., the number of replies that separate it from the main thesis
This model allows one to compute the levels in the reply tree where arguments are justified with the highest probability.

This probabilistic approach appears to be a good way to tackle the problem because it can predict the location of justified arguments in online discussions, when its results are compared to real data that we obtained from Kialo, an Internet debating platform.
The probabilistic approach also reveals three different schemes of behaviour for the probability of an argument being justified as a function of two global parameters of the reply tree: (1) the relative proportions of attacking and supporting arguments in the overall discussion
and (2) the structure of the discussion tree, as  characterised by its degree distribution.

Data from Kialo indicates that online discussions behave as trees with non-homogeneous in-degree distribution and can be classified by the proportion of supporting replies. 
When the proportion of supporting replies is  high, the proportion of justified arguments is higher in deep levels of the tree. Therefore, the ``best'' order to read the discussion comments is by starting from the deepest level (i.e. most recent comments first) and arriving to the root comment in reverse order of the level in the reply tree. 
In this case, our model suggests that a temporal ordering of new user utterances, with the most recent comments appearing first, may show a higher proportion of justified arguments than other sorting methods, but pure temporal ordering is not in itself sufficient -- it needs to be tweaked, allowing comments to be read based on the level in the reply tree rather than just the time stamp of each post. 
In contrast, when discussions are aggressive, the proportion of justified arguments is more homogeneous among the levels and there is no single ``best way'' to read the comments.

An important result that appears from our analysis is that the leaves of the discussion tree, i.e. unreplied comments,
effectively have the ``last word'', and have a great impact on the probability of all the other arguments being justified. This is due to a fundamental assumption in argumentation theory where all unrebutted arguments, and hence arguments that are not replied to, are justified by default and thereby greatly inflate the numbers of justified arguments at each level. 

It can be argued that unreplied comments may not have received sufficient scrutiny from other users, and therefore should not by themselves be counted among the justified arguments. We showed that even if  leaves were not considered in counting up ``who is justified'', the general shape of the distribution of justified arguments among the levels is still influenced by them. 
However, we also observed that in this case, when the number of attacks and supports is balanced (as in the majority of Kialo discussions), the new probability of an argument being justified per level is guided only by the number of replies to comments at that level. In an evolving discussion, the number of \emph{current} replies a comment has can therefore be an indication of its eventual inclusion among the justified arguments. 
Today's platforms support this strategy by sorting based on overall level of support (e.g. sort by number of likes and comments, or numbers of upvotes and downvotes).

A possible future improvement to our model is to depart from ``traditional" argumentation frameworks and suggest different methods to establish which arguments should be justified that dampen the high degree of influence that leaf nodes have. A possibility is giving less importance to single attacking comments, considering a node justified only if the majority of its replies are either justified and supporting the comment or unjustified and attacking it. Another possibility is to give higher importance to the judgment of arguments which have  higher number of likes or replies. In this way comments which have received a larger scrutiny have a larger influence when attacking or supporting another comment. To do this we can use for example preference-based argumentation frameworks as \cite{amgoud2002inferring}, or adapting the techniques used in \cite{DBLP:conf/comma/RagoCT16}. 

In conclusion, by characterising the locations of justified arguments in online discussions in terms of the supportiveness of the discussion ($q$) and the distribuion of leaves in the reply tree, this work points to new ways of presenting information from online discussions, or lends theoretical backing to existing methods of displaying comments in such discussions. For example, comments can be organized in a discussion such that justified arguments are presented first. Moreover comments that are particularly weak under future attacks, meaning that can easily lose their 'justified' status given their position in the graph, may be highlighted in the discussion. Our analysis can also be applied to classic ordering of comments, as by time or by likes. A work by A.P.Young \textsl{et al.}~\cite{youngranking} shows which are the sorting policies to be chosen such that more justified comments are shown first. Moreover, whatever sorting policy is used, our analysis allows to mark justified arguments to be visible by users.

To the best of the authors' knowledge, this work is the first to combine  argumentation theory with complex networks to analyse online discussions. While previous work has used argumentation theory to understand online discussions (e.g. \cite{Bosc:16,Cabrio:13}), previous research has mainly focused on \emph{mining} arguments from natural language expressions. We are motivated by the complementary question of understanding where justified arguments might be, and suggesting to the user where to look  for such justified arguments. In future work, we aim to move beyond BAFs to more sophisticated argumentation models, such as the quantitative frameworks described in \cite{leite2011social}.

One criticism that can be leveled is that normatively justified arguments do not represent "true" justified arguments because 
users may not be convinced by justified comments that do not support their point of view. 
There has been research relating how people perceive how arguments disagree and when they are justified (e.g. \cite{cramer2018directionality, cramer2019empirical, cramer2019scf2,rahwan2010behavioral}), showing that user preferences matter. On the one hand, we argue that our ``skeptical'' approach of only accepting justified arguments is the ``most suitable'' approach in situations of low trust, such as large-scale discussions. On the other hand, if UI designers are to decide on an order of presentation of comments based on our results, it is important that users find those orderings useful and convincing.
We plan on exploring such a problem in our future work by conducting user studies and experiments.

\begin{acks}
G. Boschi, C. Cammarota and N. Sastry acknowledge funding from the Engineering and Physical Sciences Research Council  (EPSRC)  through  the  Centre  for  Doctoral  Training in  Cross  Disciplinary Approaches to Non-Equilibrium Systems (CANES, Grant Nr.\  EP/L015854/1). N. Sastry and A. P. Young acknowledge funding from the Space for Sharing (S4S) project (Grant No.\ ES/M00354X/1).
S. Joglekar is supported through King's India scholarship offered by the KCL Centre for Doctoral Studies.
\end{acks}

\bibliographystyle{ACM-Reference-Format}
\bibliography{sample-base}
\end{document}